\documentclass{aastex63}

\usepackage{float}
\usepackage{amsmath}
\usepackage{comment}
\usepackage{graphicx}
\usepackage{sidecap}
\sidecaptionvpos{figure}{c}

\usepackage{kpfonts}
\begin{document}

\title{Emergent Spectral Fluxes of Hot Jupiters: \\
an Abrupt Rise in Day Side Brightness Temperature Under Strong Irradiation}

\shorttitle{Spectral Fluxes of Hot Jupiters}
\shortauthors{Deming et al.}

\correspondingauthor{Drake Deming}
\email{ldeming@umd.edu}

\author[0000-0001-5727-4094]{Drake Deming}
\affiliation{Department of Astronomy, University of Maryland,
   College Park, MD 20742, USA}

\author[0000-0002-2338-476X]{Michael~R.~Line}
\affiliation{School of Earth \& Space Exploration, Arizona State University, Tempe, AZ 85287, USA}

\author[0000-0002-5375-4725]{Heather~A.~Knutson}
\affiliation{Division of Geological and Planetary Sciences,
California Institute of Technology, Pasadena, CA 91125, USA}

\author[0000-0002-1835-1891]{Ian~J.~M.~Crossfield}
\affiliation{Department of Physics and Astronomy, University of Kansas, Lawrence, KS 66045, USA}

\author[0000-0002-1337-9051]{Eliza~M.-R.~Kempton}
\affiliation{Department of Astronomy, University of Maryland,
   College Park, MD 20742, USA}

\author[0000-0002-9258-5311]{Thaddeus~D.~Komacek}
\affiliation{Department of Astronomy, University of Maryland,
   College Park, MD 20742, USA}

\author[0000-0003-0354-0187]{Nicole~L.~Wallack}
\affiliation{Division of Geological and Planetary Sciences,
California Institute of Technology, Pasadena, CA 91125, USA}

\author[0000-0002-3263-2251]{Guangwei Fu}
\affiliation{Department of Astronomy, University of Maryland,
   College Park, MD 20742, USA}
\affiliation{Present address: Department of Physics and Astronomy, Johns Hopkins University, Baltimore, MD 21218 USA}

\begin{abstract}
We study the emergent spectral fluxes of transiting hot Jupiters, using secondary eclipses from {\it Spitzer}.  To achieve a large and uniform sample, we have re-analyzed all secondary eclipses for all hot Jupiters observed by {\it Spitzer} at 3.6- and/or 4.5\,$\mu$m. Our sample comprises 457 eclipses of 122 planets, including eclipses of 13 planets not previously published.  We use these eclipse depths to calculate the spectral fluxes emergent from the exoplanetary atmospheres, and thereby infer temperature and spectral properties of hot Jupiters. We find that an abrupt rise in brightness temperature, similar to a phase change, occurs on the day side atmospheres of the population at an equilibrium temperature between 1714\,K and 1818\,K (99-percent confidence limits). The amplitude of the rise is $291\pm49$\,Kelvins, and two viable causes are the onset of magnetic drag that inhibits longitudinal heat redistribution, and/or the rapid dissipation of day side clouds.  We also study hot Jupiter spectral properties with respect to metallicity and temperature inversions.  Models exhibiting 4.5\,$\mu$m emission from temperature inversions reproduce our fluxes statistically for the hottest planets, but the transition to emission is gradual, not abrupt. The {\it Spitzer} fluxes are sensitive to metallicity for planets cooler than $\sim 1200$\,Kelvins, and most of the hot Jupiter population falls between model tracks having solar to 30X-solar metallicity.      
\end{abstract}


\section{Introduction}
 
The Spitzer Space Telescope leaves a rich legacy of transiting exoplanetary science. {\it Spitzer} investigators obtained photometry at secondary eclipse for many transiting planets \citep{Deming_2020}. Using {\it Spitzer's} photometry, atmospheric properties such as metallicity, C/O, and temperature structure, have been inferred for dozens of hot Jupiters (e.g., \citealp{Harrington_2007, Charbonneau_2008, Knutson_2008, Fressin_2010, Madhu_2011, Todorov_2013, Beatty_2014, Mansfield_2018}), albeit not without some debate \citep{Hansen_2014, Ingalls_2016} and some revisions \citep{Knutson_2012, Diamond-Lowe_2014, Line_2016}.  The physical properties of hot Jupiters were recently reviewed by \citet{Fortney_2021}.  

Most of {\it Spitzer's} exoplanetary eclipse observations comprise photometry in broad spectral bands. Atmospheric inferences using photometry can be problematic because photometric spectral resolution is insufficient to identify molecular absorptions in emergent spectra via band shapes.  Moreover, if hot Jupiters vary widely in their properties, then studies of individual planets may struggle to accurately infer average properties of the entire hot Jupiter population.  One solution to these problems is to focus on population studies of relatively large samples of planets. Population studies of emergent fluxes have been pursued using eclipses in the {\it Spitzer} data (\citealp{Wallack_2019}, \citealp{Garhart_2020} - hereafter G20, \citealp{Baxter_2020} - hereafter B20, \citealp{Wallack_2021} - hereafter W21, and \citealp{Goyal_2021}), and also using eclipses observed with HST \citep{Mansfield_2021}.

\subsection{Motivation for This Work}

Although analyses of {\it Spitzer's} secondary eclipses have been extensive, there are 13 hot Jupiters (listed below) whose eclipses have not yet been published.  In some cases, hot Jupiters with published eclipse analyses have additional unanalyzed eclipses in the archive. Moreover, there is arguably value to re-analyzing all of the eclipses using one uniform method. To that end, we have re-analyzed all of {\it Spitzer's} secondary eclipse data in the 3.6- and 4.5\,$\mu$m bands using one method. These two bandpasses encompass the vast majority of {\it Spitzer's} secondary eclipse measurements of hot Jupiters.

We seek to establish a population-level context for observations of hot Jupiters by JWST, and by high-resolution ground-based spectroscopy using the ELTs.  The context that we seek is to establish the major properties of the hot Jupiter population that can be deduced from their emergent spectral fluxes as observed by {\it Spitzer}.  Emergent fluxes are basic observable properties of the planets, and we aim for a simple and robust analysis. 

\subsection{Organization of This Paper}

Section~\ref{sec: archival} describes the broad strategy of our archival analysis. The sample of planets that we analyze is summarized in Section~\ref{sec: sample}, and our technique for re-analyzing {\it Spitzer's} secondary eclipse data to produce emergent fluxes is described broadly in Section~\ref{sec: derivation}, and in detail in the Appendix. We interpret the emergent fluxes using a grid of models that are described in Section~\ref{sec: models}. The magnitudes and observed properties of the emergent fluxes are described in Section~\ref{sec: emergent}, and the spectral properties by comparing fluxes versus wavelength are explored in Section~\ref{sec: spectral_properties}.  Section~\ref{sec: discussion} discusses the atmospheric implications of our results, beginning with an acknowledgement of previous work in Section~\ref{Sec: Noticed}.  The potential causes of an abrupt increase in day side flux are explored in Sections~\ref{sec: weaker} to Section~\ref{sec: other}.  Section~\ref{sec: metallicities} describes the implications for atmospheric metallicities, and spectral emission from temperature inversions is discussed in Section~\ref{sec: structure}. Section~\ref{sec: summary} summarizes our results, and Section~\ref{sec: future} notes the need for future work.

\section{An Archival Analysis} \label{sec: archival}

    \subsection{The Sample of Planets}\label{sec: sample}
Our sample comprises 122 hot Jupiters, and we have re-analyzed every eclipse observed for those planets using {\it Spitzer} at 3.6- and/or 4.5\,$\mu$m.  That includes eclipses that occur during phase curve observations. Planets whose {\it Spitzer} eclipses were not previously published include: WASP-11b, -21b, -28b, -32b, -35b, -37b, -38b, -50b, -95b, -107b, -122b, HAT-P-34b, and HD\,202772b.  In total, we derive eclipse depths for 439 individual eclipses, and we perform some averaging for 18 additional eclipses of four planets.  Our derived eclipse depths are listed in Table~\ref{table:depths}.  We use the eclipse depths to derive emergent day side fluxes, and brightness temperatures ($T_b$), listed in Table~\ref{table:fluxes}.  The source of stellar and planetary parameters used in our analysis (TEPCat) is ever-evolving.  Hence, we list the stellar and planetary radii and temperatures that we adopted in Table~\ref{table: parameters}. In addition to our re-analyzed eclipses, our analysis in this paper makes use of {\it Spitzer} eclipses observed at 8\,$\mu$m, but we have not re-analyzed those eclipses, nor have we listed their emergent fluxes in Table~\ref{table:fluxes}.  Our analysis also considers eclipses from 2\,$\mu$m ground-based observations reported in the literature (i.e., \citealp{croll_2015}, and many others).       
    
    \subsection{Derivation of Eclipse Depths, Emergent Fluxes, and Brightness Temperatures}\label{sec: derivation}
    
We derive eclipse depths using the method described by G20, in fact using the same code.  This method separates systematic signals due to the IRAC instrument from the eclipse signal using pixel-level-decorrelation (PLD, \citealp{Deming_2015}).  Some minor differences from the G20 methodology are as follows.  First, to increase the practicality of our analysis, we perform aperture photometry on the {\it Spitzer} images using only apertures of fixed radius (e.g., not variable radii as per \citealp{Lewis_2013}), and centering the apertures using a 2-D Gaussian fit to the stellar image (i.e., not using flux-weighted centering, \citealp{Piskorz_2018}). This reduces the volume of photometry by a factor of four, and G20 found that fixed radii apertures with Gaussian centroiding was their best method, and was consistently good for all of the planets in their sample.  G20 also derived two depth solutions for each eclipse: one depth that maximized the fit to the Allan deviation relation for the photometric residuals (see their Section~3.4), and another depth derived from the centroid of the posterior distribution of eclipse depth. They preferred the former method in their analysis (but they commented that the population-level results were not sensitive to the choice). Using our larger sample, we find that the centroid of the posterior distribution is preferable, but we have also verified that the conclusions of this paper would not change if we used the alternate choice of eclipse depth.  There are other (minor) technical issues that arise in our analysis, and we include a more comprehensive technical discussion in the Appendix.

We believe that our results are an improvement over previous Spitzer population studies in 5 ways:  1) We have added 13 additional planets to the sample, 2) We analyze the maximum number of eclipses per planet (previous population studies have sometimes relied on literature results that do not include all of the eclipses per planet, e.g., see the discussion of HD\,189733b in Section C of the Appendix).  3) Our results are homogenous, using the same method to extract all of the eclipse depths (whereas the literature is inhomogenous, with the early Spitzer papers using much more simplistic data analysis methods), 4) a systematic comparison of seven data analysis techniques \citep{Ingalls_2016} concluded that our PLD method was among the three most accurate methods, and had the least bias of the seven.  Finally, 5) we have consulted recent high spatial resolution imaging results to revisit dilution corrections needed for eclipses where those corrections were not previously applied (Appendix Section~\ref{sec: dilution}).
 
 For the host stars, we adopt stellar effective temperatures, radii, and surface gravities from TEPCat\footnote{https://www.astro.keele.ac.uk/jkt/tepcat/}, and also planetary equilibrium temperatures, radii, and orbital distances from TEPCat \citep{Southworth_2011}.  We interpolate in a grid of ATLAS9 models \citep{Castelli_2003} to determine the fluxes of the host stars, integrating over the bandpass sensitivity functions of {\it Spitzer's} IRAC instrument.  We derive an emergent flux at the top of the exoplanetary atmosphere, using the measured eclipse depth and the stellar fluxes from the ATLAS models. To derive planetary brightness temperatures, we calculate eclipse depths for a grid of blackbody planets, integrating over the IRAC bandpasses, and we interpolate the measured eclipse depth in that grid to find the brightness temperature.  To calculate uncertainties, we propagate the uncertainty of the observed eclipse depths through the process to produce error bars on flux and brightness temperature. The planetary fluxes and brightness temperatures are given as averages per planet per waveband in Table~\ref{table:fluxes}.  
 
 In some instances, especially for weak eclipses, the error bars on eclipse depth can allow negative depths.  Negative eclipse depths would formally imply negative brightness temperatures (as per some ranges in Table~\ref{table:fluxes}).  Although negative eclipse depths and brightness temperatures are not physical, it is necessary to allow them, so that population averages and trends are free of Lucy-Sweeney bias \citep{Lucy_1971}.
 
 Our error bars on $T_b$ for the planets listed in G20 are significantly less than the G20 values.  The G20 $T_b$ error bars were calculated using a Rayleigh-Jeans approximation, which is insufficiently accurate. That was a mistake by one of us (D.D.), and we suggest that the values in Table~\ref{table:fluxes} should be used in preference to the G20 error bars.  Our current error bars in the Table were calculated using the same method as B20, i.e. propagating the eclipse depth errors through the temperature dependence of the Planck function integrated over {\it Spitzer's} bandpass responses.

\section{Atmospheric Models}\label{sec: models}

We interpret our results using a set of 1D self-consistent radiative-convective-thermochemical equilibrium models. The models were previously described and used in \citet{Piskorz_2018}, \citet{Arcangeli_2018}, \citet{Mansfield_2018}, \citet{Kreidberg_2018}, \citet{gharibnezhad_2019}, \citet{Mansfield_2021}, and recently benchmarked against M-dwarfs in \citet{Iyer_2022}.  Briefly, the models solve the 1D radiative-convective equilibrium problem via a Newton-Raphson iteration to achieve zero net flux divergence across each layer interface (see \citealp{mckay_1989}).  Layer-by-layer radiative fluxes are computed using the two stream source function method from \citet{toon_1989} (without clouds/scattering) and convective flux is computed via mixing length theory.  Equilibrium chemistry with rainout (depletion of elemental species in overlying layers due to the onset of condensate formation) is computed using the NASA CEA2 routine \citep{McBride_1996}.  We include $\sim 40$ line and continuum opacity sources accounting for species relevant in chemical equilibrium from 300 - 4000\,Kelvins covering 0.1 - 100\,$\mu$m (UV species included are neutral atomics/ions, H$_2$, CO, and SiO), with molecular species sourced from the latest ExoMol \citep{Tennyson_2020} line lists and neutral/ion atomic species from the Kurucz line-list database.  Line opacities are computed using the methods described in \citet{gharibnezhad_2021} and then are converted into R=250 correlated-K coefficients (using a 10 point double-Gauss quadrature sampling for the bin weights/ordinates, e.g., \citealp{molliere_2015}) with mixed opacities computed on-the-fly using the RORR method described in \citet{amundsen_2017}. 

The models are in two groups: first, we calculate model tracks using a planet of 1-Jupiter mass and radius, irradiated by a solar-type star at a range of orbital distances. For these tracks, we adopt solar abundances in the exoplanetary atmosphere, and either complete re-distribution of stellar irradiance, or no re-distribution (these are same models as in \citealp{Mansfield_2021}). We found the no-redistribution case to be especially useful in our analyses, and we henceforth refer to that case as the no redistribution (NR) track.  Our second group of models is constructed for individual planets, using their measured masses and radii, and the temperatures and radii of their host stars; these are “population”-level models. In all cases, we use PHOENIX model atmospheres to specify the irradiating spectrum from the star. We checked that using PHOENIX stellar models to irradiate the planets, versus ATLAS9 stellar models to calculate emergent fluxes (Section~\ref{sec: derivation}) did not produce significant inconsistencies - see Section D of the Appendix.

\begin{SCfigure}
\includegraphics[width=3in]{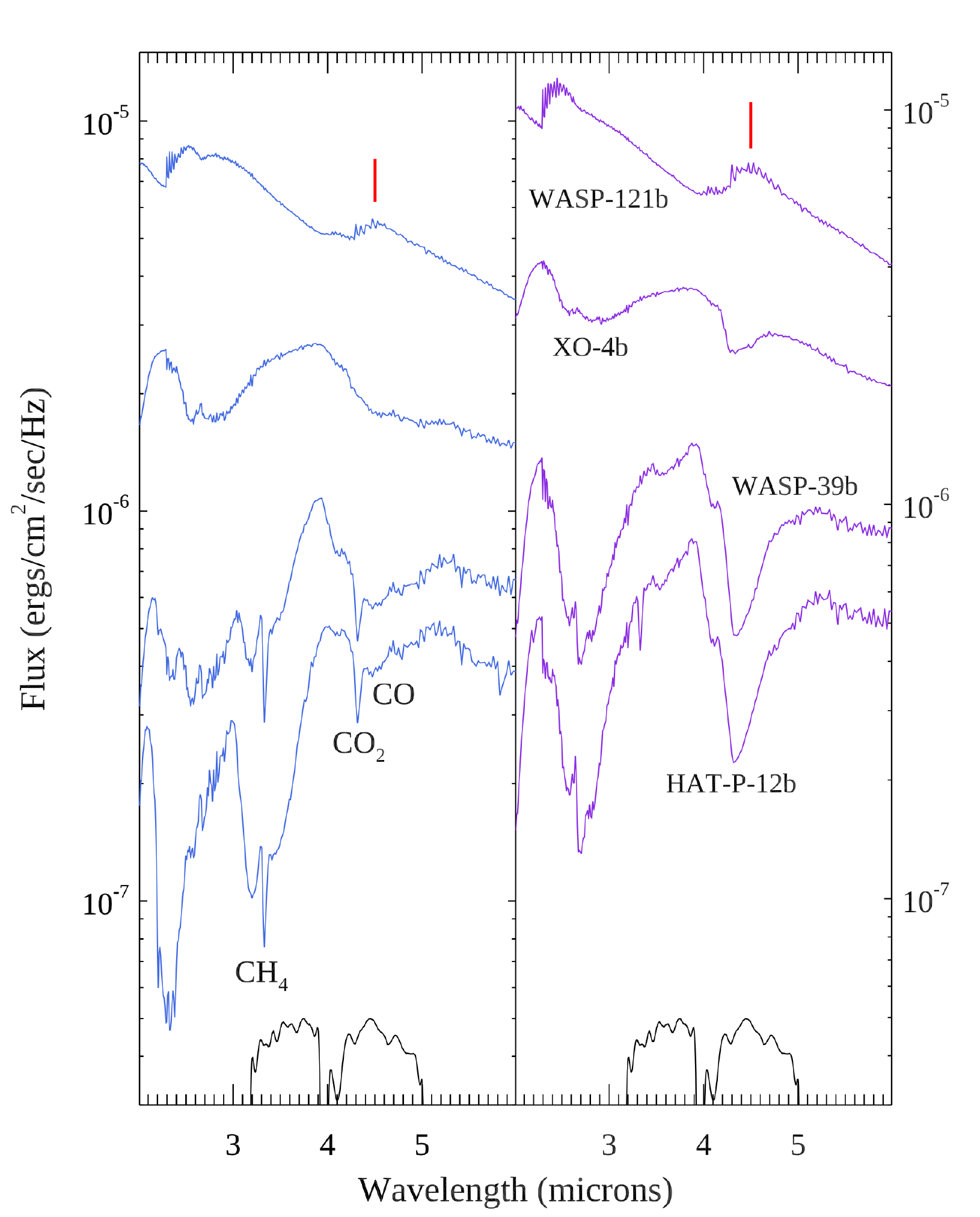}
\caption{Examples of modeled emergent flux for planets with a range of temperature, and in the wavelength region encompassing {\it Spitzer's} 3.6- and 4.5\,$\mu$m bandpasses (response curves at the bottom of the panels).  The left panel shows models having solar metallicity, and the right panel is 30-times solar metallicity for the same planets.  The planets and equilibrium temperatures are: WASP-121b ($T_{eq}=2358$\,K), XO-4b (1641\,K), WASP-39b (1166\,K), and HAT-P-12b (955\,K).  The red tic marks at 4.5\,$\mu$m call attention to carbon monoxide emission from a temperature inversion in WASP-121b.  \label{fig: several_models}}
\end{SCfigure}

The models for individual planets vary the heavy element abundance (metallicity), using both solar metallicity, and 30-times (30X) solar.  Because both groups of models omit the presence of clouds, their atmospheres are dark, with albedos near zero.  Emergent fluxes of four planets in this group of models are illustrated in Figure~\ref{fig: several_models}, that shows some important properties of the modeled fluxes. Spectral absorption features weaken as temperature increases, and carbon monoxide appears in emission due to an atmospheric temperature inversion in the hottest planets.  Increasing metallicity changes the nature of the absorption in {\it Spitzer's} bands.  At solar metallicity, methane absorption is prominent in the 3.6\,$\mu$m band for the coolest models. At high metallicity, carbon becomes preferentially bound in carbon monoxide and carbon dioxide \citep{Moses_2013}. That effect is also apparent on Figure~\ref{fig: several_models}, where the coolest 30X metallicity models exhibit absorption in the 4.5\,$\mu$m bandpass, not as much at 3.6\,$\mu$m.

\section{Emergent Fluxes and Day Side Temperature}\label{sec: emergent}

The emergent flux of a hot Jupiter in a given spectral band is determined by the vertical atmospheric temperature gradients and the specifics of the opacity sources that are significant in that band. The day side brightness temperature in a given spectral band is thereby diagnostic of the temperature structure of the atmosphere, and how it may change as a function of stellar irradiance, and how that heating is redistributed by longitudinal circulation.

Longitudinal heat circulations for individual hot Jupiters are best studied using thermal infrared phase curves, and those studies have been extensive (e.g., \citealp{Knutson_2007, Knutson_2012, Wong_2015, Stevenson_2017, Kreidberg_2018, Bell_2021, May_2021, May_2022, Dang_2022}). Phase curves can delineate emergent flux over the full range of planetary longitude, whereas eclipses yield only the day side flux. Nevertheless, it has long been known that the day side fluxes can yield important statistical information on the nature of heat circulation for hot Jupiters \citep{Cowan_2011}.  Although phase curves are numerous, eclipse data are even more numerous. Eclipse fluxes provide an opportunity to probe physical effects on the day side atmosphere with maximum resolution in stellar irradiance as we now demonstrate.

\begin{SCfigure}
\centering
\includegraphics[width=3in]{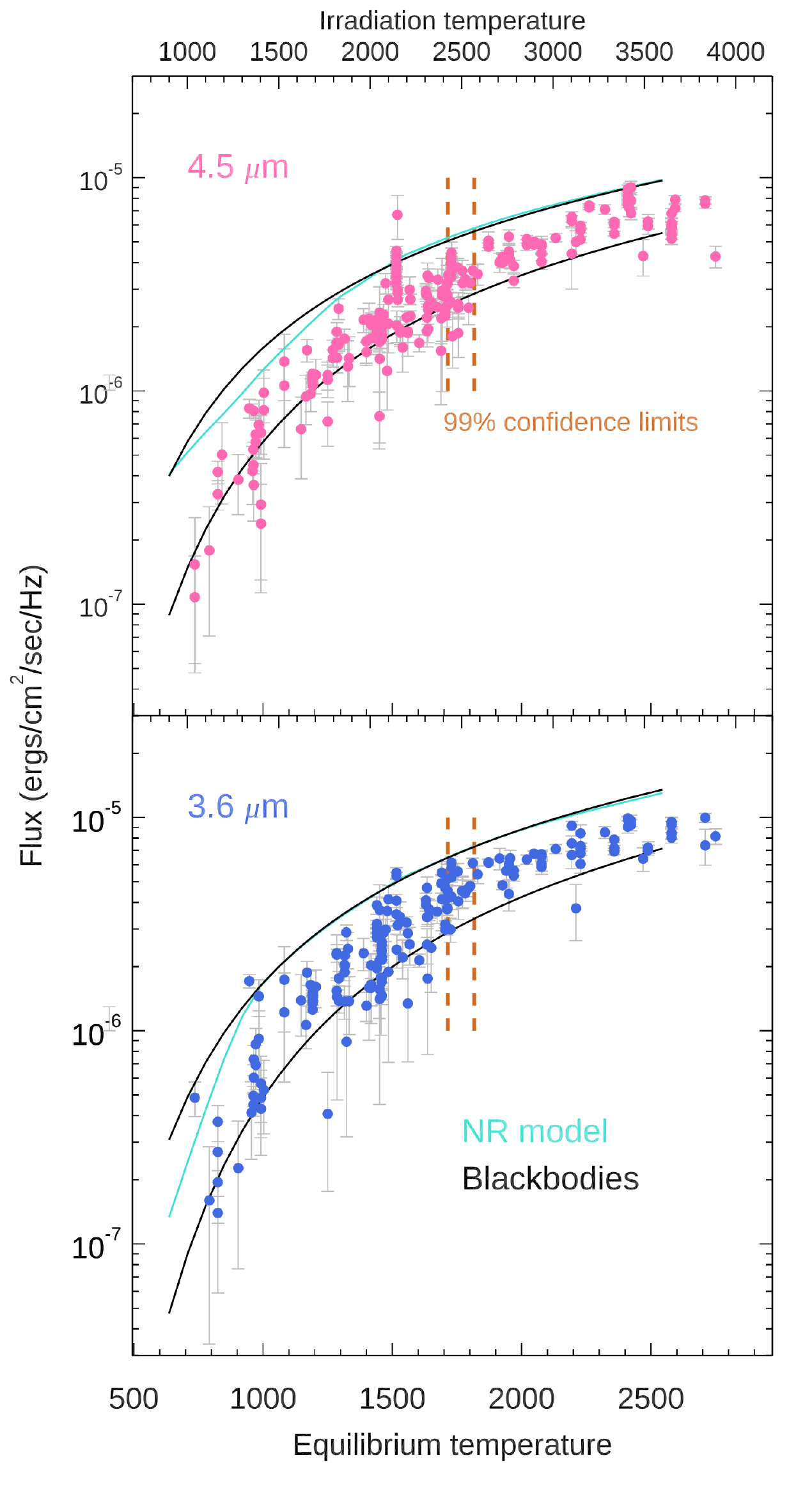}
\caption{Observed and modeled fluxes for hot Jupiters at secondary eclipse, in both the {\it Spitzer} 3.6- and 4.5\,$\mu$m bands, versus planetary $T_{eq}$ and $T_{irr}$. Fluxes from the no redistribution (NR) model are also plotted as a track, as well as the tracks for two blackbodies.  The upper blackbody adopts zero albedo and no heat redistribution, whereas the lower blackbody adopts zero albedo and heat redistribution over the entire planet. The vertical dashed lines mark the 99\% confidence regions for the quasi-discontinuous increase in flux (see text for discussion). \label{fig: flux34}}
\end{SCfigure}

Figure~\ref{fig: flux34} shows the emergent spectral fluxes that we derive for hot Jupiters at both 3.6- and 4.5\,$\mu$m, plotted as a function of the equilibrium and irradiation temperatures, $T_{irr}.$ \footnote{irradiation temperature is $T_{irr} = {T_s}\sqrt{R_s/a}$, where $T_s$ and $R_s$ are the stellar temperature and radius, and $a$ is the orbital distance of the planet in a circular orbit.}  Note that analyzing fluxes at each wavelength allows us to make maximum use of planets observed in only one {\it Spitzer} bandpass (e.g., 4.5\,$\mu$m, program 14059). 

For equilibrium temperature ($T_{eq}$), we adopt an albedo of zero and complete redistribution of heat. Equilibrium temperature is less than irradiation temperature by a factor of $\sqrt{2}$. Our values refer to the emergent flux density at the top of the exoplanetary atmosphere, so they are independent of the planetary radius. We derive uncertainties on the fluxes by propagating uncertainties in the eclipse depth, stellar temperature, and planetary radius.  Uncertainty in the planetary radius enters only via the ratio of planetary-to-stellar radius. That ratio contributes negligibly to the flux uncertainty because transit depths are known much more precisely than eclipse depths.  We similarly derive uncertainties in the equilibrium and irradiation temperatures based on uncertainties in the stellar temperature and planetary orbit.   

\begin{SCfigure}
\centering
\includegraphics[width=3in]{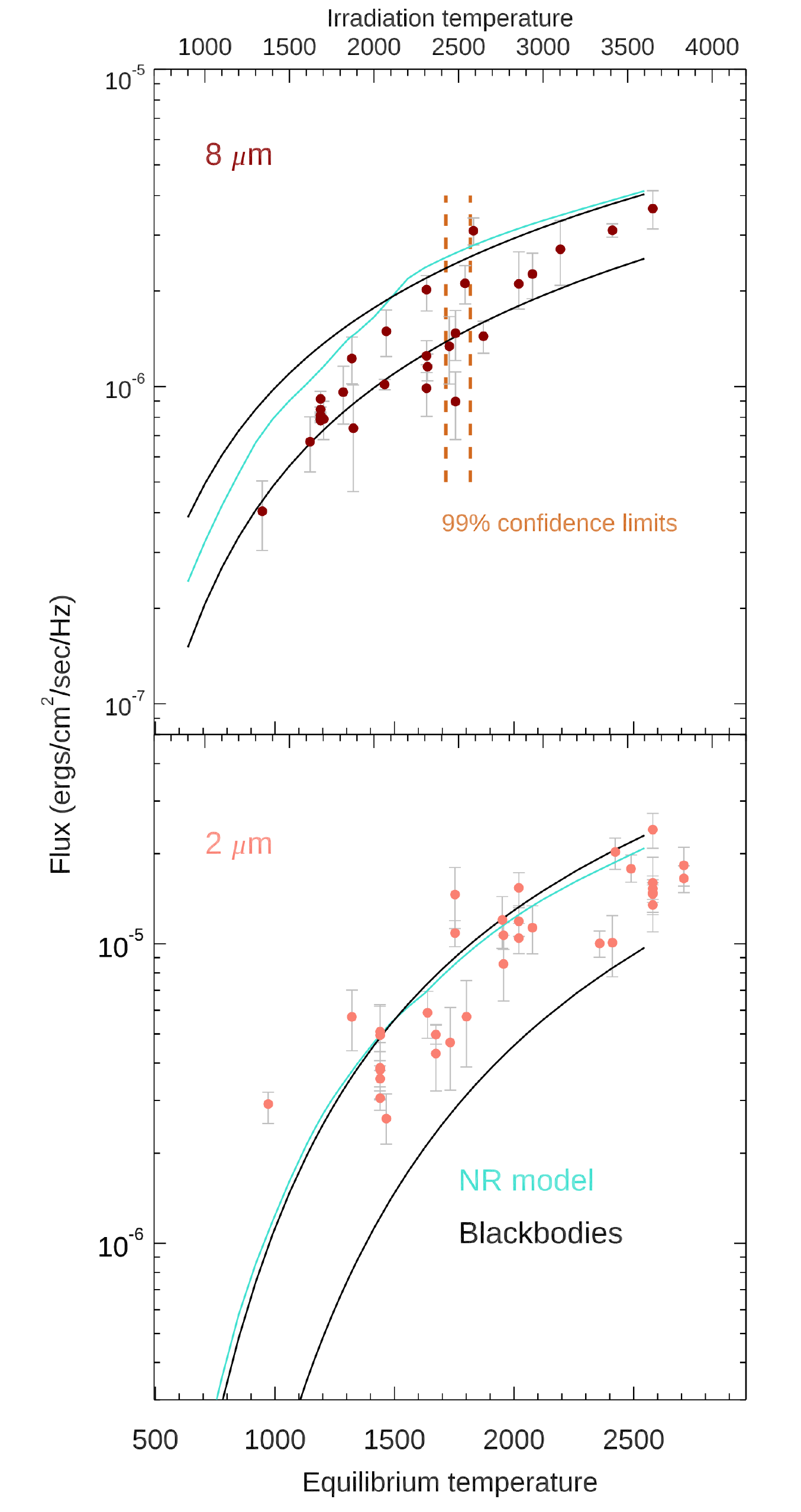}
\caption{Observed and modeled fluxes for hot Jupiters at secondary eclipse, in both the {\it Spitzer} 8\,$\mu$m bands, and ground-based eclipses in the K-band (2.3\,$\mu$m), versus planetary $T_{eq}$ and $T_{irr}$. No redistribution (NR) model and blackbody tracks are overplotted as per Figure~\ref{fig: flux34}. The vertical dashed lines mark the 99\% confidence regions for the quasi-discontinuous increase in flux (see text for discussion). \label{fig: flux28}}
\end{SCfigure}

Fluxes from the no redistribution (NR) models, and two blackbodies, are overplotted on Figure~\ref{fig: flux34}.  The NR model track adopts no redistribution of stellar irradiance, solar composition, and surface gravity equal to Jupiter's.  Temperature via absorbed irradiation is the overwhelmingly dominant factor that determines the flux in the model track.  We experimented with plotting models tailored to the parameters inferred for individual planets (i.e., surface gravity, composition), and thereby verified that those parameters have a minor (albeit detectable) effect in this context. The blackbody tracks on Figure~\ref{fig: flux34} correspond to zero albedo and no heat redistribution, and also heat redistribution over the entire planet. 

Figure~\ref{fig: flux28} shows fluxes that we derive from literature values of eclipses observed using the next two most widely-used bands, i.e. {\it Spitzer's} 8\,$\mu$m band, and ground-based eclipses in the K-band (e.g., \citealp{Kovacs_2019}, and many others), and also the NR model track and blackbody tracks. 

Both Figures~\ref{fig: flux34} and \ref{fig: flux28} exhibit similar behavior in the {\it Spitzer} bands.  For $T_{eq}$ exceeding $\sim 1700$\,K, the planetary fluxes are largely confined between the two blackbody curves. Below $T_{eq}\sim 1700$\,K, fluxes can fall below the lower blackbody curve. This general behavior has been seen previously, for example by G20 (their Figure~16), and \citet{Goyal_2021}.  Fluxes below the lower blackbody curve require either molecular absorption, or a non-zero albedo (e.g., reflecting clouds) that lower the atmospheric temperature below the complete redistribution case, or cold clouds that emit little thermal radiation.  The Figures suggest that the transition from fluxes that drop below the lower blackbody, to fluxes that remain above it, occurs rather abruptly near $T_{eq} \sim 1700$\,K.  It has long been known that the hottest of the hot Jupiters circulate heat with the least efficiency, and thereby have the hottest day side temperatures \citep{Cowan_2011, Schwartz_2015, Showman_2020}. Below, we define this rise in emergent flux level quantitatively, using the largest possible collection of {\it Spitzer's} secondary eclipses.  

\subsection{An Abrupt Rise in Flux}\label{sec: rise}

We here demonstrate that an abrupt rise (i.e., nearly discontinuous versus $T_{eq}$) in day side emergent flux occurs at an equilibrium temperature between $1714$\,K to $1818$\,K.  In this section, we establish the data-related aspects of the rise, such as amplitude and statistical significance.  In Section~\ref{sec: discussion} we explore the physical implications for the day side atmosphere. 

First, we note that the 2\,$\mu$m fluxes behave differently than the {\it Spitzer} fluxes. They occur more often above the NR model, which is physically difficult to explain. Ground-based eclipse observations are difficult, and it seems possible that they may be especially affected by selection bias.  Observers may be reluctant to publish eclipses that give weak detections (and consequently low fluxes), but they may readily publish strong eclipses. Hence, we hypothesize that the ground-based eclipses may suffer from selection bias.  In order to further investigate the abrupt rise in emergent flux, we combine the three {\it Spitzer} bands, but for now we omit the 2\,$\mu$m band.  Figure~\ref{fig: shift} plots the ratio of the emergent flux in the {\it Spitzer} bands to the value from the NR model track, versus $T_{eq}$, with each point being a single eclipse. The NR model track is smooth (Figure~\ref{fig: flux34} and Figure~\ref{fig: flux28}), so this division removes the gradual increase in flux with increasing temperature, but it does not create the abrupt rise in flux: that abrupt rise is caused by the observed data, not by the models.  

\subsubsection{Partial Eclipses and Eccentric Planets}\label{sec: eccentric}

Figure~\ref{fig: shift} omits the planets WASP-34b \citep{smalley_2011}, WASP-67b \citep{hellier_2012}, and WASP-140b \citep{hellier_2017}, because their eclipses are partial, and therefore the eclipse depth cannot be converted reliably to an emergent flux at the top of their atmospheres.  We do include TrES-3b, because \citet{odonovan_2007} concluded that it is not partial, although it is close.  Consequently, we include it but we increase its error bars by calculating the fraction of flux that would be missed when varying the orbital parameters within the error ranges given by \citet{odonovan_2007}.  

Planets with eccentric orbits may have day side temperatures at eclipse that are not representative of their equilibrium temperature averaged over their orbits.  This primarily affects HD\,80606b and HAT-P-2b, due to their substantial orbital eccentricities \citep{moutou_2009, bakos_2007}.  For these two planets, we replaced their TEPCat values of $T_{eq}$ with the day side temperatures calculated at eclipse in the time-dependent models of \citet{mayorga_2021} (see Table~\ref{table: parameters}). Planets with less extreme eccentricities (e.g., WASP-14b, \citealp{joshi_2009}) are included using $T_{eq}$ values from TEPCat.

\subsubsection{Amplitude and Location of the Flux Rise}\label{sec:amplitude}

In order to investigate the magnitude and location of the emergent flux rise, we apply the Bayesian methodology described by \citet{Lee_1977} (hereafter, LH) to the ratio data of Figure~\ref{fig: shift}. The LH method is specifically tailored to find abrupt changes in the values of serial data.  It adopts a uniform prior for the change location, and assumes that the data values and change amplitude have Gaussian distributions.  It gives the posterior distributions for the change magnitude and location as analytic expressions involving sums over the data values. (The posterior distributions are not Gaussian, or even necessarily smooth.)  

The LH algorithm detects an abrupt rise in the ratio of the emergent flux to the NR model. The distribution for temperature where the rise in flux ratio occurs is sharply peaked at $T_{eq} =1730$\,K, as shown on Figure~\ref{fig: shift}.  The rise in flux is abrupt, i.e. not clearly resolved in the temperature coordinate in spite of using 437 eclipses.  The posterior distribution for the rise amplitude has a mean value of $0.12\pm0.0157$ in the ratio of flux to NR model (from $0.58$ to $0.70$), so the rise is nominally detected at the $7.6\sigma$ level of significance, assuming that the data are normally distributed.  As a test, we impose a strong (and physically motivated) prior on the flux ratio by using only the flux ratios per eclipse that are between 0 and 1.  That range would encompass low fluxes (e.g., due to high, cold clouds), ranging to high fluxes from a hot day side with no redistribution, and assuming a blackbody spectrum. In that case, the location and amplitude of the flux rise are not significantly changed, but the statistical significance would rise to $9\sigma$, principally because ratios above unity are eliminated. 

Although we do not include the 2\,$\mu$m eclipses in our adopted solution, we explored whether including them would affect the result. In this case, we reject only the eclipse of WASP-10b at 2\,$\mu$m, that is an extreme outlier \citep{cruz_2015}, at 2.65 times the NR model.  Including the balance of the 2\,$\mu$m eclipses in the LH analysis does not significantly change the location of the flux rise, and the statistical significance actually rises to $7.8\sigma$.  Although it is informative that the 2\,$\mu$m eclipses support our result from the {\it Spitzer} eclipses, we continue to base our analysis exclusively on the {\it Spitzer} data.

\subsubsection{Three Tests}\label{sec: tests}

\begin{figure*}[t]
\centering
\includegraphics[width=7in]{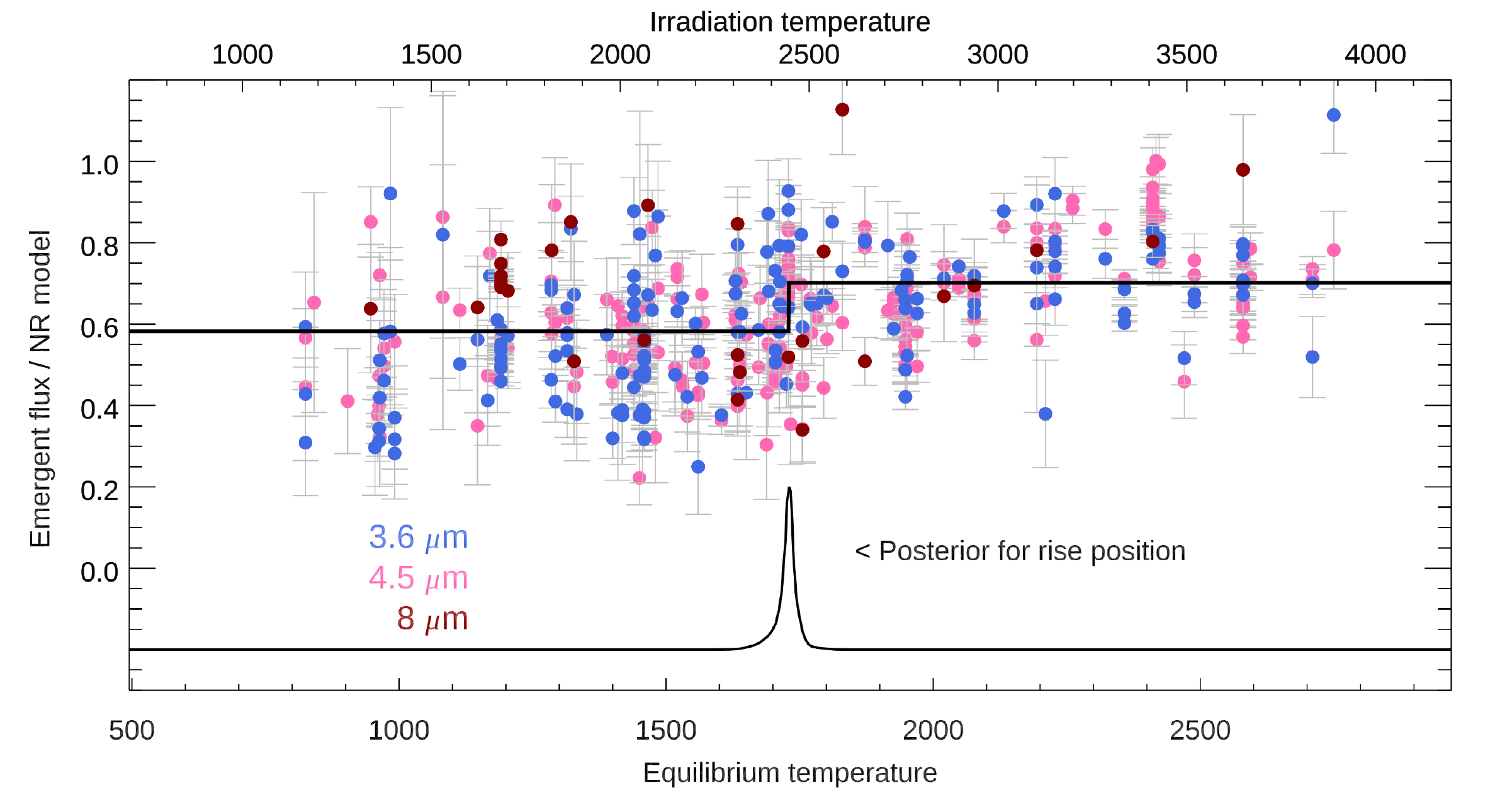}
\caption{Ratio of emergent flux to the flux from the no redistribution (NR) model, versus equilibrium and irradiation temperature. Each of the 437 points is a single {\it Spitzer} eclipse, except for a few planets (WASP-29, HAT-P-12, and HAT-P-15) where we averaged several eclipses due to low signal-to-noise ratio.  Several outlying points lie above the range of this plot, but they have been included in the analysis. The fluxes exhibit an abrupt rise at an equilibrium temperature exceeding 1730$\,$K, and the horizontal lines mark the inferred levels (ratios of 0.58 and 0.70). The amplitude of the rise corresponds to a brightness temperature increase of $210\pm22$\,K, consistent among the different wavelength bands.  The lower trace is the posterior distribution for the rise position, based on the analytic Bayesian methodology of \citet{Lee_1977}. The horizontal scale of the posterior distribution for the solid line has been expanded graphically by a factor of 5, to make the shape more visible. For visual clarity, we do not plot error bars on the X-axis (temperature), but those uncertainties (median value $\pm19$\,Kelvins) were included when deriving the posterior distribution for the temperature of the flux rise.}
\label{fig: shift}
\end{figure*}

We did three additional tests to verify the existence of an abrupt rise in flux on Figure~\ref{fig: shift}.  To verify the statistical significance of the rise, we made $10^6$ synthetic realizations of the flux ratio data.  Those realizations preserved the $T_{eq}$ values and the relative differences among multiple eclipses for the same planet, but they varied the average flux ratios for each planet.  We adopted a uniform (not Gaussian) distribution for the flux ratios, and distributed them between 0 and 1 (see above), with no trends or rises in average value.  Because we adopted a {\it uniform} distribution for these flux ratios, not a Gaussian distribution assumed by the LH method, these synthetic data represent a "worst case" for the rise analysis, assuming that there is no physical reason to justify ratios above unity (i.e., no planets hotter than the no redistribution case). Only 0.028\% of these simulations produced a detected rise at a significance equal to the real data and we conclude that the flux rise is significant at greater than a 99.97\% level of confidence.

Our second test explored whether the flux rise might be gradual, rather than abrupt. We again constructed $10^6$ realizations of the data, preserving the $T_{eq}$ values and relative differences among eclipses of the same planet.  We scattered the average flux ratio for each planet around a gradual ramp with a scatter matching the real data.  The ramp varied between the two flux ratio levels detected in the real data, and spanning the range between 1000\,K and 2500\,K in $T_{eq}$. The purpose of these realizations is to investigate whether the LH algorithm could be fooled into mis-identifying a gradual flux rise as being abrupt.  We found that only 0.009\% of the simulations were mis-identified as being abrupt. We also did a third test, where we again made $10^6$ realizations of the data that scattered around a slope consistent with the real data, but with no restriction to vary between two flux ratio levels. This test also found only 0.009\% of the simulations were mis-identified as being abrupt. Hence, we trust the abrupt nature of the flux rise inferred from the LH algorithm. We also conducted further tests of the LH algorithm, as described below.

\begin{SCfigure}
\includegraphics[width=3.5in]{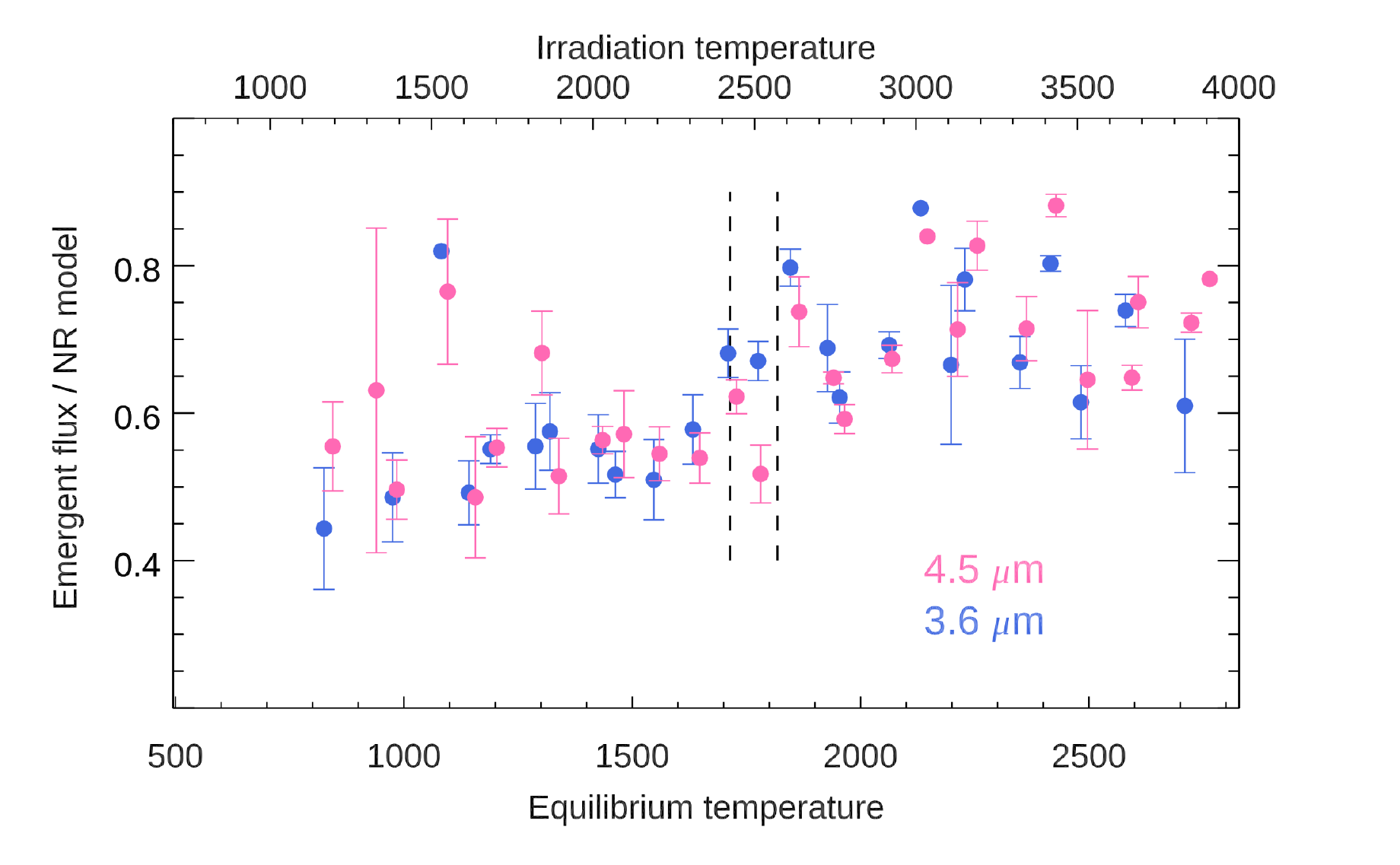}
\caption{3.6- and 4.5\,$\mu$m points from Figure~\ref{fig: shift}, binned over intervals of 100\,Kelvins, with error bars based on the standard deviation (scatter) within each bin, divided by the square root of the number of points in that bin (standard error of the mean). The bin centers are nominally at multiples of 100K in equilibrium temperature, but the points deviate slightly in X because we average both the X- and Y-values within each bin.  We plot only the 3.6- and 4.5\,$\mu$m points so as to maximize the clarity of the abrupt rise in flux.  The vertical dashed lines are the 99\% confidence limits that we infer for the temperature of the abrupt rise in flux. Note also that the data plotted here were not used directly in our analysis - this Figure is only to make the flux rise optimally visible to the eye. }\label{fig: ratio_binned}
\end{SCfigure}

Figure~\ref{fig: shift} contains many points, and it may be difficult to see the abrupt flux rise when so many points are plotted.  To make the flux rise visually clearer, Figure~\ref{fig: ratio_binned} bins the 3.6- and 4.5\,$\mu$m points from Figure~\ref{fig: shift} in arbitrary intervals of 100\,Kelvins in equilibrium temperature. The formulation of the data on Figure~\ref{fig: ratio_binned} is not used directly in our analysis; that Figure is intended merely to make the abrupt rise in flux more clear to the eye. 

\subsection{Location of the Abrupt Rise in Flux}

The LH algorithm has been widely used in terrestrial meteorology (e.g., to detect rapid changes in weather patterns), but it is not often used in astronomy.  We further tested the methodology using more Monte Carlo simulations applied to both the real data as well as synthetic data.  We verified that, as the transition in flux level is broadened in temperature, the LH posterior distribution also broadens.  We explored the effect of dithering the data in both X- and Y-coordinates, using the observational errors in both observed flux ratio and temperature.  We found that the observational errors in temperature are not sufficient to significantly broaden the posterior distribution in rise location, and the effect of the errors in temperature is included on Figure~\ref{fig: shift}.

The observed planets are not distributed uniformly in temperature, and we hypothesized that the sharpness of the LH posterior distribution in flux-rise temperature could be an indirect artifact of the temperature distribution of the planet population. We therefore perturbed the real data, forcing the flux rise at $T_{eq} = 1730$\,K to occur over broader ranges. We broadened a range of the data near $1730$\,K over intervals of 50, 100, 150, 200, 250, and 300\,K.  Broadening the data means that we adjusted the flux ratios (but not the equilibrium temperatures) to scatter around an average linear ramp within those intervals. As we increased the width of the ramp, the $1730$\,K peak in the posterior distribution decreased in amplitude (as expected), and a second peak became evident near $1800$\,K. 

Based on all the numerical experiments described above, we conclude that the flux rise at $T_{eq} = 1730$\,K is statistically robust and abrupt, and not a random or gradual fluctuation.  The sharp peak in the posterior distribution at $1730$\,K accurately reflects the specific distribution of equilibrium temperature and emergent flux in our actual data.  However, we also conclude that small and statistically plausible variations in our data are consistent with the flux rise occurring near $1800$\,K.  We compiled the $T_{eq}$ values that mark the 1-percent and 99-percent limits in the combined area of both peaks of the posterior distribution.  That conservative prescription gives us 99\% confidence that the flux rise occurs between equilibrium temperatures of $1714$\,K and $1818$\,K.  That $104$\,K range is sufficiently small to indicate a rapid change in exoplanetary atmospheres, similar to a phase change.

\subsection{Temperature Amplitude of the Rise}\label{sec: tempamp}

The analysis above focused on the rise in the ratio of emergent flux to the NR model's flux.  That ratio rises because the emergent fluxes from the planets increase with increasing $T_{eq}$, not because the fluxes from the models drop.  However, it is more physically meaningful to express the rise in flux as an increase in the planetary brightness temperature, $T_b$. We calculate the temperature rise using two methods.  First, we convert the rise in the flux ratio average at each wavelength (Figure~\ref{fig: shift}) to flux units using the NR model fluxes, and then calculating the rise in $T_b$ by reference to the temperature dependence of the model flux. Using this method, the rise in $T_b$ is $\Delta{T_b}=192\pm34$\,K at 3.6\,$\mu$m, $\Delta{T_b}=216\pm30$\,K at 4.5\,$\mu$m, and $\Delta{T_b}=249\pm69$\,K at 8\,$\mu$m. Within the errors, the rise in brightness temperature is the same at all three {\it Spitzer} wavelengths. Weighting the temperature rise at each wavelength by $\sigma^{-2}$ yields an average temperature rise of $\Delta{T_b}=210\pm22$\,K.  

\begin{figure}
\centering
\includegraphics[width=4in]{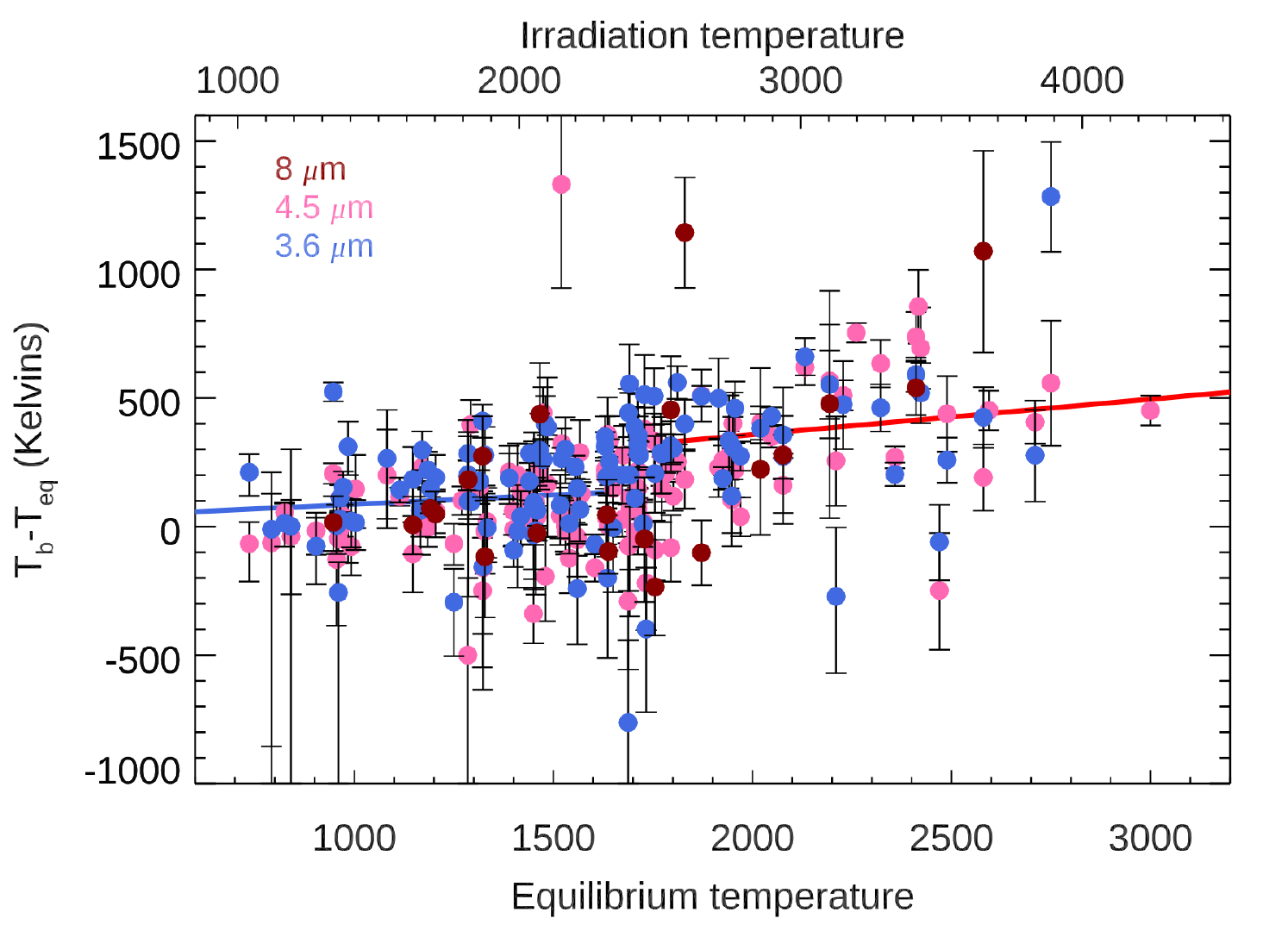}
\caption{Brightness temperature difference ($T_b$ minus $T_{eq}$) versus equilibrium temperature.  Each point represents the average of all eclipses for a given planet in a given wavelength band (see color legend).  The location of the abrupt temperature rise is consistent with $T_{eq} = 1730$K (as on Figure~\ref{fig: shift}), with amplitude  $\Delta{T_b}=291\pm49$\,K (see text).  The red and blue lines are least-squares fits in the regimes hotter and cooler than the location of the abrupt temperature rise. The fits consider uncertainties in both the X- and Y-coordinates, but for visual clarity the (small) error bars on X are not plotted.  The upward slopes of the lines are consistent with brightness temperature exceeding the equilibrium temperature by increasing amounts with increasing irradiance.  \label{fig: Trise}}
\end{figure}

A second method to calculate the rise in $T_b$ is illustrated in Figure~\ref{fig: Trise}. In this case, we calculate $T_b$ for each planet at each wavelength using the eclipse depths and the {\it stellar} model fluxes.  $T_b$ is generally proportional to the equilibrium temperature, so we plot $T_b - T_{eq}$ in order to remove the overall increase in temperature with increasing irradiance, and concentrate on the fine structure in the proportionality.  Because the most strongly irradiated planets circulate heat with the least efficiency, we expect that $T_b - T_{eq}$ will slope upward (albeit, slightly) as a function of $T_{eq}$.  We applied the LH algorithm to the Y-axis of Figure~\ref{fig: Trise}, and it finds a discontinuity near $T_{eq}=1800$\,K, with $6.7\sigma$ significance.

As a further test, we solved for a linear slope via least-squares for the entire range of data on Figure~\ref{fig: Trise}, and a second solution that uses two separate least-squares lines. One of those two lines is for $T_{eq}$ less than the discontinuity (blue line on Figure~\ref{fig: Trise}), and a second line at temperatures above the discontinuity (red line on Figure~\ref{fig: Trise}). Based on the $\chi^2$ values, we calculate Bayesian Information Criteria for the one line versus two line solution, and we find $\Delta{BIC}=14$. That value of $\Delta{BIC}$ indicates "very strong" evidence \citep{Kass_1995} in favor of the two line solution, consistent with the LH algorithm finding a discontinuity in both the emergent flux (Figure~\ref{fig: shift}) and brightness temperature (Figure~\ref{fig: Trise}).

Each point on Figure~\ref{fig: Trise} is a single wavelength band for a single planet, which would over-weight planets that were observed in multiple bands. To derive the most accurate value of the temperature rise at the discontinuity, we weight each {\it planet} in the population equally.  That yields a rise in brightness temperature of $\Delta{T_b}=291\pm49$\,K, consistent (at $1.7\sigma$) with the calculations based on the emergent fluxes (Section~\ref{sec: tempamp}, above).

To summarize our brightness temperature analysis, we find that $T_b$ increases as a function of equilibrium temperature with a slope exceeding unity, but that the relation has a discontinuity in the range of $T_{eq}$ between $1714$ and $1818$\,Kelvins.  The existence of the discontinuity in brightness temperature is detected by the LH algorithm \citep{Lee_1977}, and supported very strongly by a Bayesian Information Criterion applied to results from least-squares.  The amplitude of the discontinuity in brightness temperature ($\Delta{T_b}=291\pm49$\,K) is (within the errors) equal at 3.6-, 4.5- and 8\,$\mu$m.

We discuss the interpretation and atmospheric implications of the temperature rise in Section~\ref{sec: discussion}.

\section{Spectral Properties of the Emergent Fluxes}\label{sec: spectral_properties}

We now turn to more general spectral properties of the emergent fluxes and brightness temperatures that are given in Table~\ref{table:fluxes}.  Previous investigations have focused on this topic.  Because eclipses of hot Jupiters have been most abundantly observed in {\it Spitzer's} 3.6- and 4.5\,$\mu$m bands, the comparison between those bands has been a specific focus of population studies (e.g. G20, B20, W21, and \citealp{Goyal_2021}, and also \citealp{Dransfield_2020}). We here use our increased sample size to revisit the 3.6- and 4.5\,$\mu$m flux comparison, and also to construct a color-flux diagram.

\subsection{3.6- Versus 4.5 Micron Brightness Temperatures}

\begin{figure*}
\centering
\includegraphics[width=6in]{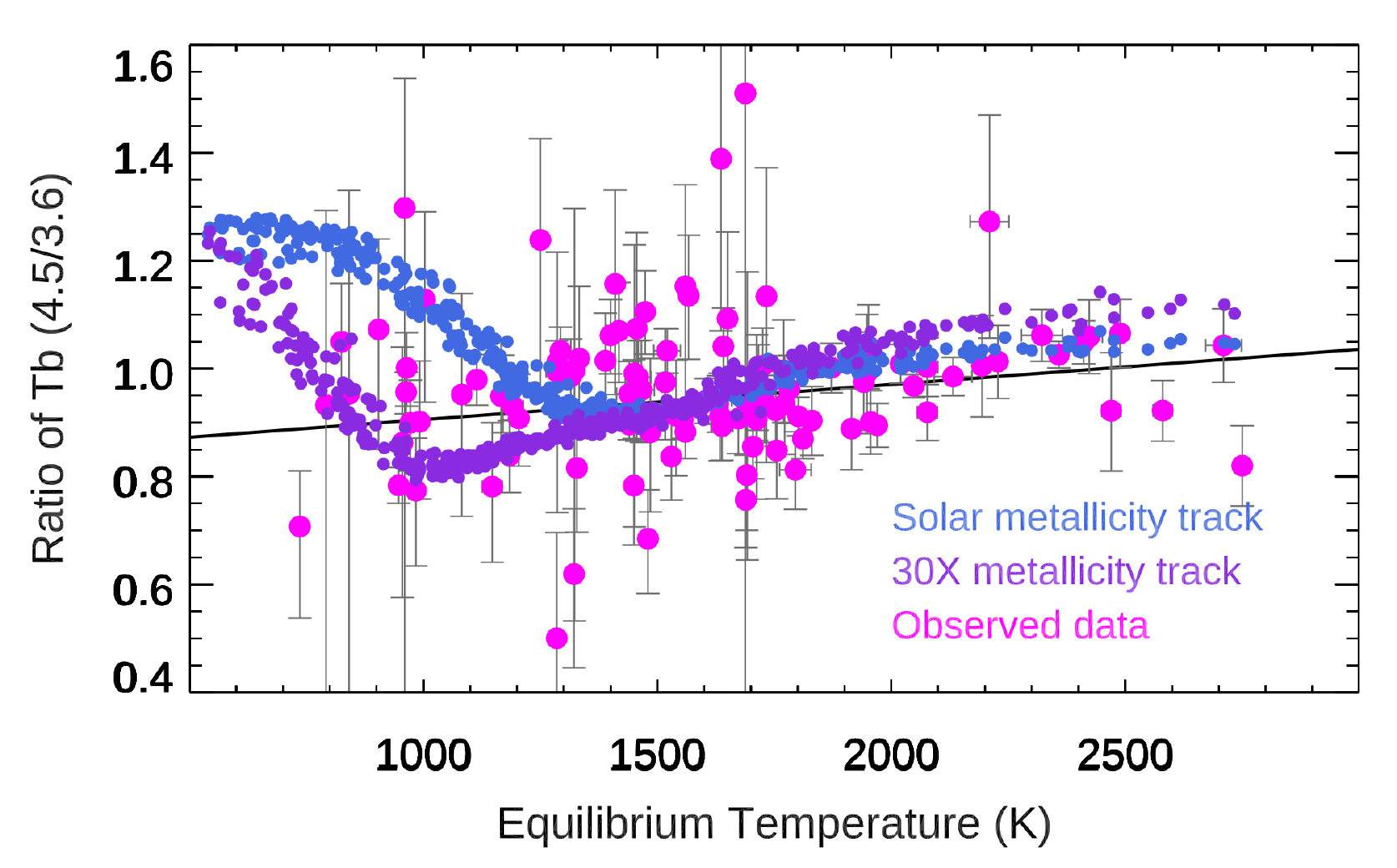}
\caption{Brightness temperature ratio ($T_b$ at 4.5\,$\mu$m divided by $T_b$ at 3.6\,$\mu$m) versus equilibrium temperature.  The tracks of points in slate blue and blue-violet colors are theoretical points from the models (solar metallicity, and 30-times solar metallicity). The black line is a least-squares fit that considers error bars in both the X- and Y-coordinates (error bars in X are not plotted, to keep the Figure clear). The slope of the fitted line differs significantly from zero (see text), and the model tracks come close to bracketing the observed points for equilibrium temperatures below $\sim 1200$\,Kelvins.}
\label{fig: Tb_slope}
\end{figure*}

Figure~\ref{fig: Tb_slope} shows the ratio of 4.5- to 3.6\,$\mu$m brightness temperature, versus equilibrium temperature.  G20 found that this ratio increases with equilibrium temperature with a slope of $100 \pm 24$ parts-per-million per Kelvin over the range of equilibrium temperature from approximately 800 to 2500 Kelvins.  Work with larger samples (B20, and W21) found a similar trend, and our complete sample of {\it Spitzer's} hot Jupiters confirms the effect.  A least-squares fit of a straight line to the data on Figure~\ref{fig: Tb_slope}, and accounting for uncertainties of each point in both X and Y coordinates, yields a positive slope of $67\pm13$ ppm per Kelvin.  The slope is thereby significantly different from zero by $5.0\sigma$, revealing a spectral property of the hot Jupiter population. Recall that G20 - using a smaller sample - found a slope of $100\pm24$ ppm per Kelvin, so our result is $1.4\sigma$ below G20's estimate and thereby consistent with their work.  B20 found essentially the same result as G20, and W21 found a similar relation (albeit expressed differently), at $3.2\sigma$ significance.  

We also allow for the fact that the scatter on Figure~\ref{fig: Tb_slope} exceeds the error bars, for example due to intrinsic variation in the population. We scaled the error bars to greater values so that the error bars statistically match the total scatter in the points (observational errors plus intrinsic variations).  In that case, the least-squares solution gives a slope of $65\pm17$ ppm per Kelvin, still exceeding zero by $3.7\sigma$.  We did another test, following W21.  We made $10^5$ synthetic data sets whose per-point scatter was identical to the real data, but no systematic trend was present.  We calculated least-squares slopes for those synthetic data realizations.  Because the synthetic data sets contain no trend, least-squares slopes as large as in the real data can only occur via random chance.  If the number of those cases is very small, then we can conclude that the real data contain a real slope to appropriately high confidence.  We find that only 0.30\% of the realizations produce a slope as large as we find in the real data.  We conclude that the real data exhibit a significant slope at 99.7\% confidence. The brightness temperature at 4.5\,$\mu$m increases relative to the 3.6\,$\mu$m brightness temperature as the planets get hotter. The planets do not behave as blackbodies, because blackbodies have the same brightness temperature at all wavelengths (ratio everywhere equal to unity on Figure~\ref{fig: Tb_slope}).

We calculated 3.6- and 4.5\,$\mu$m brightness temperatures for solar and 30X metallicity model tracks, and those tracks are also plotted on Figure~\ref{fig: Tb_slope}.  For equilibrium temperatures exceeding $\sim 1200$\,Kelvins, the models slope upward similarly to the data, so the models do a good job of accounting for the hottest planets (the solar metallicity track is best for the hottest planets).  However, for equilibrium temperatures below $\sim 1200$\,Kelvins, the data and models differ widely, and we discuss the physical implications of this difference in Section~\ref{sec: metallicities}. 

  \subsection{A Color-Flux Diagram}\label{sec: color_flux}

By analogy with stellar population studies, color-color and color-magnitude diagrams have been used to infer properties of the hot Jupiter population \citep{Triaud_2014a, Triaud_2014b, Dransfield_2020}. We explore that approach using our complete sample, but with one key difference.  Rather than plotting a color-magnitude diagram, we plot a color-flux diagram wherein the radii of the planets is not a factor, only the spectral character of their emergent flux matters. 

\begin{figure*}
\centering
\includegraphics[width=7in]{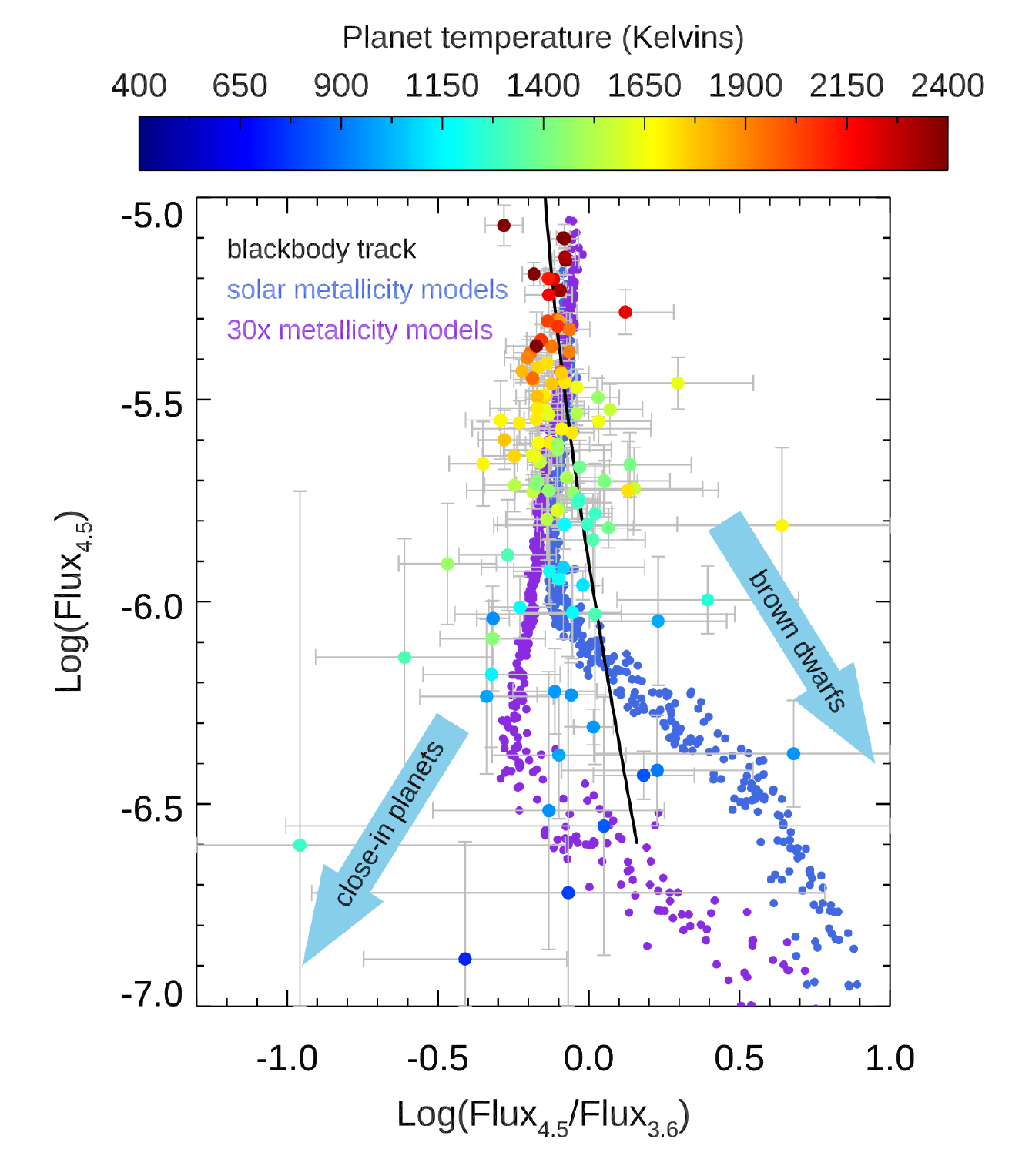}
\caption{Color-flux diagram for hot Jupiters.  The X-axis is log of the flux ratio (4.5\,$\mu$m flux divided by 3.6\,$\mu$m flux) and the Y-axis the log of the 4.5\,$\mu$m flux. The black line is a blackbody track, and the tracks for two collections of model planets are also included: solar metallicity models, and 30X solar metallicity. The observed  planets are color-coded by calculated equilibrium temperature.}  \label{fig: HR}
\end{figure*}

Figure~\ref{fig: HR} shows the color-flux diagram for our complete sample, using the 4.5\,$\mu$m flux on the Y-axis, and with the planets color-coded by equilibrium temperature.  We also plot two model tracks, based on collections of models for individual planets (Section~\ref{sec: models}).  Those tracks are for solar metallicity, and for 30X solar metallicity.  The hottest planets tend to emit the most flux, and thereby lie highest on the Y-axis.  For those hottest planets, the model tracks are very close together, indicating that the flux ratio (4.5\,$\mu$m to 3.6\,$\mu$m) is not very sensitive to metallicity, in agreement with Figure~\ref{fig: Tb_slope}.  The models track the hottest planets very well, unlike the blackbody line that slopes slightly to the upper left. \citet{Dransfield_2020} found that planets tend to scatter to the right of the model track in a color-magnitude diagram (their Figure~4); we find good agreement with the models for the hottest planets.  Also, Figure~\ref{fig: HR} is consistent with Figure~\ref{fig: Tb_slope}, wherein the brightness temperature ratio (4.5- to 3.6\,$\mu$m) increases with equilibrium temperature (blackbodies having a constant ratio), and the models follow the planets in Figure~\ref{fig: Tb_slope} for equilibrium temperatures exceeding $\sim 1200$\,K. 

One aspect of the hottest planets deserves mention, i.e. that there is no obvious and abrupt shift toward greater 4.5\,$\mu$m flux relative to 3.6\,$\mu$m for the hottest planets (i.e. no abrupt shift to the right on Figure~\ref{fig: HR}).  Such a shift was found by B20, that they attributed to excess emission in the fundamental carbon monoxide band caused by atmospheric temperature inversions.  We discuss this issue in more depth in Section~\ref{sec: structure}.

The coolest planets in the sample fall to the lower left on Figure~\ref{fig: HR}, unlike directly imaged planets and brown dwarfs that fall to the lower right \citep{Dransfield_2020}. The error bars are large for individual planets, but the shift of the hot Jupiter population to the lower left is a real effect. This effect is consistent with previous work (e.g., \citealp{kammer_2015}), and is also consistent with the behavior seen on Figure~\ref{fig: Tb_slope}.  Much of this effect can be due to high metallicity, because the 30X solar model track follows the planets on Figure~\ref{fig: HR} much better than does the solar metallicity track.  We discuss this issue in greater depth in Section~\ref{sec: metallicities}.  

\section{Discussion}\label{sec: discussion}

We here discuss that the flux rise has been noticed in previous work, we explore the physical implications of the abrupt flux rise, and we also discuss the spectral properties of the emergent fluxes. 

\subsection{Noticed Previously}\label{Sec: Noticed}

The flux rise is noticeable on the right panel of Figure~2 in \citet{Wallack_2021}, and was also noted by \citet{Goyal_2021}. The latter authors did not elaborate, but \citet{Wallack_2021} discussed decreasing efficiency of heat transport for temperatures exceeding $\sim 1500$\,K, albeit they did not emphasize the abruptness of the transition in their paper. We believe that this is the same phenomenon that we measure in this paper, in spite of the modest difference in the location of the transition.  B20 found an abrupt increase in 4.5\,$\mu$m flux near $T_{eq}=1660\pm100$\,K that we believe is related to the same phenomenon, as we discuss in Section~\ref{sec: 
 structure}.

\subsection{Weakening Molecular Absorption}\label{sec: weaker}

The first question concerning the abrupt flux rise is whether it reflects an increase in the temperature of the exoplanetary atmosphere, or whether it is merely an apparent increase due to a change in the depth of formation.  For example, if molecular absorption weakens, then the emergent flux can originate from deeper and hotter layers of the atmosphere.  That could cause an increase in emergent flux without any change in the run of temperature versus height in the atmosphere.  The depth of flux formation could change due to either molecular dissociation or cloud dissipation.  

The models show that molecular absorption does indeed weaken as the equilibrium temperature increases, but the weakening is gradual, not abrupt.  Moreover, the abrupt increase in brightness temperature being closely equal at 3.6- and 4.5\,$\mu$m is unlikely to be caused by weakening of molecular absorption.  The molecular opacities are different in the two bands, and the 4.5\,$\mu$m band includes emission by carbon monoxide, not present at 3.6\,$\mu$m. We conclude that weakening molecular absorption is not a likely explanation for the abrupt rise in flux. 

\subsection{Water Vapor Dissociation}\label{sec: cooling}

The abrupt flux rise occurs near an equilibrium temperature where \citet{Mansfield_2021} find that water absorption disappears in emergent spectra due to the dissociation of water vapor \citep{Parmentier_2018}. Water vapor is a cooling agent: stellar irradiance is absorbed primarily at optical and near-IR wavelengths, and water vapor helps to balance that heating by re-emitting in the infrared. Therefore the absence of water vapor can produce a heating of the atmosphere.  To the extent that water dissociation occurs abruptly with rising equilibrium temperature, it could in principle produce the abrupt flux rise that we detect.  However, this explanation also fails.  Figure~\ref{fig: shift} shows the flux rise as an increase in observed flux {\it divided by the NR model's flux}.  The NR model {\it includes} the effects of water vapor and its dissociation on the atmospheric temperature structure.  Therefore the abrupt flux rise seen in Figure~\ref{fig: shift} cannot be ascribed to lack of cooling by water vapor.   

\subsection{Cloud Clearing}\label{sec: clearing}

Cloud dissipation that abruptly reveals deeper, hotter layers, is a possible explanation.  Current understanding of cloud formation and dissipation \citep{Helling_2019} predicts that the dominant cloud types change with temperature.  The day sides of the ultra-hot Jupiters are expected to be cloud-free \citep{Parmentier_2016, Wakeford_2017, Helling_2021}. However, near the temperature of the flux rise, multiple types of clouds are expected to be present \citep{Adams_2022}. Could cloud-clearing occur over a small range of temperature and thereby lead to the abrupt flux increase?  Or, does cloud-clearing occur more gradually in temperature, as clouds of a single composition are replaced by other cloud types as temperature increases? Shifts in Kepler phase curves \citep{Demory_2013} are diagnostic of the presence of reflecting clouds \citep{Parmentier_2016}. Those phase shifts change gradually with equilibrium temperature (e.g., Figure 3 of \citealp{Parmentier_2016}), suggesting that cloud clearing is not sufficiently abrupt to account for the flux rise that we detect. The trend of geometric albedo versus equilibrium temperature for hot and ultra-hot Jupiters \citep{Wong_2021} also seems inconsistent with a single rapid cloud-clearing process.  

Nevertheless, some models indicate that rapid cloud dissipation could indeed produce the abrupt increase in brightness temperature that we observe. \citet{Roman_2019} and \citet{Roman_2021} modeled cloud formation in GCM models of hot Jupiters. Figure~2 of \citet{Roman_2021} shows an abrupt increase in day side thermal emission between $1750$\,K and $2000$\,K for their extended nucleation-limited case.  Also, Figure 2 of \citet{Gao_2021} shows a distinct jog upward in modeled day side brightness temperature near $1700$\,K, having approximately the amplitude that we observe and caused by the dissipation of day side silicate clouds.  

We conclude that cloud dissipation is a tentatively viable explanation for the abrupt increase in brightness temperature that we observe.  However, additional work is needed in order to clarify the effects of cloud dissipation on the day side brightness temperature, and to reconcile observations at multiple wavelengths.

\subsection{Efficiency of Heat Transport}\label{sec: efficiency}

It is also important to consider a rise in day side temperature due to an abrupt decrease in the efficiency of spatially redistributing stellar irradiance.  \citet{Parmentier_2021} found that non-gray radiative transfer enables a decrease in heat redistribution efficiency that begins suddenly near $1600$\,K, and continues to increase monotonically with increasing irradiance. Our data indicate a similar behavior but require a more abrupt decrease in heat redistribution concentrated near $1730$\,K. We discuss whether that abrupt decrease in efficiency could occur either as a thermodynamic effect of night side clouds, or via inhibition of winds via magnetic drag.  

\subsubsection{Thermodynamic Effect of Night Side Clouds}

Hot Jupiters having tidally locked rotation can balance stellar irradiation by re-emitting that energy input, especially from their night sides. However, cloud formation on the night side reduces the efficiency with which the night side can radiate, leading
to increased day-night contrasts \citet{Parmentier_2021}. As equilibrium temperatures rise, night side clouds are forced to higher altitudes \citep{Gao_2021}, but the temperature of the clouds tends to be approximately constant, determined by the condensation temperature of the cloud species. That can further decrease the efficiency of re-radiation by the night side, forcing the day side to become hotter and re-radiate more strongly in order to balance stellar irradiation.  That effect is intertwined with the consequences of cloud-clearing on the day sides, where the dissipation of day side clouds allows deeper and hotter layers of the day side atmosphere to become visible.  

\subsubsection{Magnetic Drag}\label{sec: drag}

\citet{Perna_2010} and \citet{Menou_2012} calculated that rising ionization in strongly irradiated hot Jupiters would produce a conductive atmosphere and magnetic induction and drag that damps the zonal winds, leading to a hotter day side.  For a reasonable atmospheric field strength of 10 Gauss, Menou's Figure~1 indicates that the zonal winds fall in amplitude by half near an equilibrium temperature of $\sim 1650$\,Kelvins.  That is close to where we observe the abrupt rise in day side flux, and Menou's scaling calculations predict that the decrease in zonal wind speed becomes complete over $\sim 200$\,Kelvins in equilibrium temperature. 
 
Self-consistent 3D MHD simulations \citep{Rogers_2014b} also show a similar decrease in zonal wind (their Figure~6) at $T_{eq} \sim 1500$ to $1800$\,K, depending (like Menou's predictions) on the assumed magnetic field strength. \citet{Hindle_2021} also find a transition in MHD behavior as ionization sets in and the magnetic Reynolds number becomes large. 
 
\citet{Thorngren_2018} infer the efficiency of stellar irradiation at inflating the radii of 281 giant planets via Ohmic dissipation. They find a peak in efficiency near the equilibrium temperature of our flux rise, and \citet{Sarkis_2021} confirmed their results with an independent study. Inflation of radii by Ohmic dissipation is intimately tied to the presence of magnetic drag \citep{Menou_2012, Rauscher_2012, Rauscher_2013}, and \citet{Knierim_2022} applied the results from \citet{Thorngren_2018} to an Ohmic dissipation model. They found that interpreting hot Jupiter radius inflation as due to magnetism implies a strong decrease in zonal wind speed at $\sim 1700$ to $1800$\,K, consistent with our abrupt increase in day side brightness temperature.
 
We conclude that the onset of magnetic drag joins possible cloud dissipation as viable explanations for the abrupt rise that we detect in day side emergent flux and brightness temperature.  Section~\ref{sec: future} discusses some tests that could help to distinguish between these possibilities, (or demonstrate that they both play a role).

\subsection{Other Factors in the Flux Rise}\label{sec: other}

We have also considered other factors that can contribute to the abrupt flux rise.  The hottest planets tend to orbit the hottest stars.  We applied the LH algorithm to the trend of stellar host temperature as a function of planetary equilibrium temperature, but we find no abrupt change in stellar temperature corresponding to the flux rise.  However, indirect factors related to the stellar temperature are possible contributors.  For example, the spectral distribution of stellar energy can enhance temperature inversions \citep{Lothringer_2019}, and potentially play a role in the flux rise.  Other factors such as chemical disequilibrium, metallicity, and variations in C/O ratio (Section \ref{sec: metallicities}) may contribute via their effect on radiative opacities. Clouds can also scatter in the infrared, and make planets on the cool side of the flux rise appear colder than they actually are \citep{Taylor_2021}.
 
 \subsection{Metallicities and C/O}\label{sec: metallicities}
 
Figures~\ref{fig: Tb_slope} and \ref{fig: HR} provide some metallicity information concerning hot Jupiters (because opacities differ between the two {\it Spitzer} bands).  For temperatures above $\sim 1200$\,Kelvins, the models track the brightness temperature ratio (Figure~\ref{fig: Tb_slope}) and color (Figure~\ref{fig: HR}) of the hot Jupiter population very well. Both of those Figures demonstrate that the planets are not blackbodies \citep{Goyal_2021}, but their departures from blackbodies are due to a combination of continuous opacity and weak molecular absorption/emission in the 3.6- and 4.5\,$\mu$m {\it Spitzer} bands, not strongly sensitive to metallicity.  Spectroscopy of individual hot Jupiters has revealed both high metallicities \citep{Sheppard_2021, Wong_2022} and sub-solar metallicites \citep{Line_2021}. The {\it Spitzer} data prefer solar metallicites for equilibrium temperatures above $\sim 1200$\,Kelvins, because the solar metallicity track on Figure~\ref{fig: Tb_slope} matches the brightness temperature ratio better than does the 30X metallicity track.  However, Figure~\ref{fig: HR} shows that many of the coolest hot Jupiters are best matched by high metallicity models, because the 30X solar track falls well to the left of the solar metallicity track. Many of the coolest planets on Figure~\ref{fig: Tb_slope} fall between the solar and 30X metallicity tracks. (We do not expect that metallicity is a function of equilibrium temperature {\it per se}, but the cooler planets more easily reveal an enhanced metallicity).

In principle, atmospheric metallicity can be a function of planetary mass and/or radius \citep{Welbanks_2019}, and that relation might be reflected in the brightness temperature ratio. We searched for a correlation between the $T_b$ ratio and both planetary mass and radius in the region where the model tracks for solar and 30X solar diverge ($T_{eq} < 1200$\,K, see Figure~\ref{fig: Tb_slope}).  We found no statistically significant correlations in either case, but only 21 planets lie in that region.  Our results remain broadly consistent with modest enhancements in the atmospheric metallicity of hot Jupiters, those enhancements being below the upper limits inferred by \citet{Thorngren_2019}.     

The C/O ratio should also be considered when comparing the relative fluxes to models \citep{Wallack_2019}. We compared the brightness temperature ratio to a model track with C/O=0.85.  That comparison (not plotted) provides only worse agreement with the data than does the solar abundance C/O track.  We conclude that enhanced C/O is not a significant help when interpreting the {\it Spitzer} eclipses.  

\subsection{Temperature Inversions and CO Emission}\label{sec: structure}

There is excellent (and growing) evidence that the most strongly irradiated planets host temperature inversions in their upper atmospheres \citep{Evans_2020, Cont_2021, Fu_2022, Yan_2022a, Yan_2022b}.  Our models for the hottest planets include temperature inversions, and emission in the v=1 to v=0 fundamental band of carbon monoxide that lies within the 4.5\,$\mu$m bandpass.  That spectral emission is a significant reason why the planets and the track of models slope to the upper right of the blackbody track on Figure~\ref{fig: HR}.  We see no evidence for a distinct transition in the population due to the emission phenomenon, as claimed by B20, but we did not look for such an effect statistically. Rather, we have focused on the abrupt increase in flux that we find to occur in both the 3.6- and 4.5\,$\mu$m spectral bands.  

B20 concluded for an abrupt transition to emission at an equilibrium temperature of $1660\pm100$\,Kelvins.  Within the errors, that location agrees with the abrupt day side brightness temperature rise that we observe. The temperature rise is easier to detect at 4.5- than at 3.6\,$\mu$m, but in our extensive and improved data we detect it in both bands. B20 matched their 3.6\,$\mu$m data with blackbodies, and then extended and removed those blackbodies from 4.5\,$\mu$m data to detect excess emission. We suggest that B20 thereby detected the day side temperature rise only at 4.5\,$\mu$m, whereas we find that the abrupt flux increase occurs in both bands.

\section{Summary and Future Work}\label{sec: summary_future}

\subsection{Summary}\label{sec: summary}

We have re-analyzed {\it Spitzer's} 3.6- and 4.5\,$\mu$m secondary eclipses of hot Jupiters using one uniform method.  Our sample comprises 457 eclipses of 122 planets. This large sample allows us to make general statements, with good statistical justification, concerning the emergent spectral properties of the hot Jupiter population.  We find that the day side brightness temperatures in the population increase abruptly, similar to a phase change, at an equilibrium temperature between $1714$ and $1818$\,Kelvins (99\% confidence interval).  The amplitude of the temperature rise is $291\pm49$\,Kelvins, and we conclude that it is caused either by rapid clearing of day side clouds, or by decreased redistribution of stellar irradiance due to magnetic drag, as predicted a decade ago by \citet{Menou_2012}.

We compare the fluxes and brightness temperatures in the 3.6- and 4.5\,$\mu$m bands, and we confirm previous work showing that the brightness temperature ratio (4.5- to 3.6) increases with equilibrium temperature.  Models having solar metallicity reproduce that brightness temperature ratio for planets with equilibrium temperatures exceeding $\sim 1200$\,Kelvins, but the {\it Spitzer} photometry is not very sensitive to metallicity for those hottest planets.  For equilibrium temperatures below $\sim 1200$\,Kelvins, the brightness temperature ratio falls consistently below the models having solar metallicity.  Most of those cooler planets are bracketed by models between solar and 30X solar metallicity.  Our results are broadly consistent with modest enhancements in the atmospheric metallicity of hot Jupiters, those enhancements being below the upper limits inferred by \citet{Thorngren_2019}.     
  
The ultra-hot portion of the population exhibit temperature inversions in our models, with emission due to carbon monoxide in the 4.5\,$\mu$m bandpass. That emission is a significant factor that allows the models to statistically reproduce the observations for the hottest planets. We do not see a distinct transition due to temperature inversions in the hot Jupiter population as deduced by \citet{Baxter_2020}, but we did not look for such an effect statistically.  Moreover, the random errors per-planet are large, and it is generally not possible to point to individual planets and conclude that they host a temperature inversion based only on the {\it Spitzer} photometry. Additional data \citep{Mansfield_2021, Fu_2022} are often needed to establish temperature inversions in individual planets.

\subsection{Future Work}\label{sec: future}

The cause of the abrupt rise in brightness temperature can be tested by future work, both theoretical and observational.  Theoretical investigations that specifically address the abruptness of the transition in the population would be illuminating.  Observationally, it is possible that the longitudinal offset in the population of phase curves could exhibit a corresponding abrupt transition to near-zero at equilibrium temperatures exceeding our range of $1714$ to $1818$\,Kelvins.  Some results in that regard are encouraging (i.e., Figure~12 of \citealp{May_2022}), but not all \citep{Bell_2021}.  It should also be possible to derive temperature-pressure profiles for hot Jupiters on both sides of the transition, using eclipse spectroscopy from JWST. Those temperature profiles may encode the effects of magnetic drag, when compared to MHD general circulation models \citep{Rogers_2014b, Beltz_2022}. The temperature structure should also encode the presence of clouds, and help to determine the role of cloud dissipation on the brightness temperatures of hot Jupiters. Finally, abrupt changes in Doppler signatures might be detectable using ground-based high resolution spectroscopy, and any such effects would be diagnostic of changes in the zonal winds and in the efficiency of heat transport.

An aspect of secondary eclipses that we have omitted from the scope of this paper concerns the orbital phase of the eclipses, and the implications for orbital eccentricity and dynamics.  Recent improvements in orbital ephemerides \citep{Ivshina_2022} imply that the secondary eclipse phases should have significant utility for studies of orbital dynamics, and we will address that topic in a future paper.

\acknowledgements

We thank Peter Gao for clarifying the signatures of cloud dissipation and for comments on this paper. We also thank the anonymous referee for comments that allowed us to improve our original version.  We acknowledge an unpublished email exchange among Heather Knutson, Jonathan Fortney, and Adam Showman, wherein they discussed a "sharp transition" in the efficiency of heat transport in hot Jupiters. They concluded that a purely radiative time scale effect would be gradual, not abrupt, and also that the onset of magnetic drag would be more likely to be gradual than abrupt. Cloud clearing was mentioned by Adam Showman as possibly causing a sharp transition, due to rapid cloud dissipation with increasing temperature. 

\clearpage
\appendix
\section{Derivation of Eclipse Depths}

\subsection{Photometry}
Extraction of a secondary eclipse depth begins with photometry of the {\it Spitzer} images, either full frame or subarray images.  For full frame images, we extract a 32x32 pixel subframe centered on the host star, and thereafter we process those data in a similar manner to the subarray data.  We first repair hot pixels  with a 4$\sigma$ filter applied to each pixel in time, and we replace discrepant pixels by their median value over time.  (Strictly speaking, discrepant pixels should be zero-weighted and not repaired, but \citet{Tamburo_2018} found that it makes negligible difference in actual practice with {\it Spitzer} photometry.)  Subarray data cubes are 32x32-pixels x 64 subframes. After the hot pixel correction, we process each of the 64 subframes in the data cubes independently. For each image we find the centroid of the star by fitting a 2-D Gaussian, after median filtering the image to assure that the Gaussian fit is not fooled by any discrepant pixels not repaired by the temporal filter.  We calculate the flux within 11 circular apertures centered on the star, the aperture radii varying from 1.6 to 3.5 pixels.  We determine the background flux by first masking the star, and then fitting to a histogram of background pixel values to determine the median.  Scaling that value to the sizes of the apertures, we subtract the background from each image independently and thereby produce a time series of photometry for each eclipse of each planet.

In the case of eclipses that occur within long time series such as phase curves, we use only a limited range of data centered on the eclipse, typically 0.4 to 0.6 in phase (when the eclipse is at phase 0.5).  Since we process 457 eclipses, we find that no rigid rules are practical, because within that large collection of eclipses are multiple special cases that frustrate any rigid procedures.  We accordingly sometimes vary the range of data that we extract from phase curves depending on factors such as the amplitude of the phase variation and any instrumental temporal ramps that may be present.  In several instances, we checked to make sure that the derived eclipse depths do not vary systematically with the range of the data that are used. For eclipses observed independently of phase curves, we omit the first 30 minutes of the data, to avoid temporal ramps that are often steepest at the start of the time series.  That 30 minute trim does not include "peak up" data that are often acquired before starting the primary time series observations.  In some cases, when a temporal ramp is especially prominent, we omit 60-, or (in very few cases) 90 minutes of data at the start of the time series.  

\subsection{Pixel-Level Decorrelation}
{\it Spitzer} data are affected by a well known intra-pixel sensitivity variation that must be removed from the data in order to derive accurate eclipse depths.  We do that using pixel-level decorrelation (PLD, e.g., G20, \citealp{Deming_2015, Wong_2016, Kilpatrick_2017, Benneke_2017}), based on 12 pixels centered on the star (a 4x4 square, minus the corners).   The PLD fitting process solves simultaneously for the eclipse depth, central phase, and temporal ramp coefficients.  The analysis is formally Bayesian, but the nature of secondary eclipses give us either very weak prior constraints, or very strong ones, so in practice the solution is dominated by minimizing chi-squared.  For example, the central phase of the eclipse is only weakly constrained by transit and radial velocity data, so we specify a uniform interval for the prior on central phase.  Conversely, the shape of the eclipse curve is very strongly constrained by transit and RV data.  Hence we use priors with zero variance for orbital parameters other than central phase (i.e., we freeze their values).  This is consistent with previous work (G20, and \citealp{ORourke_2014, Evans_2015, Mansfield_2018, Deming_2019}). 

\begin{SCfigure}
\centering
\includegraphics[width=3in]{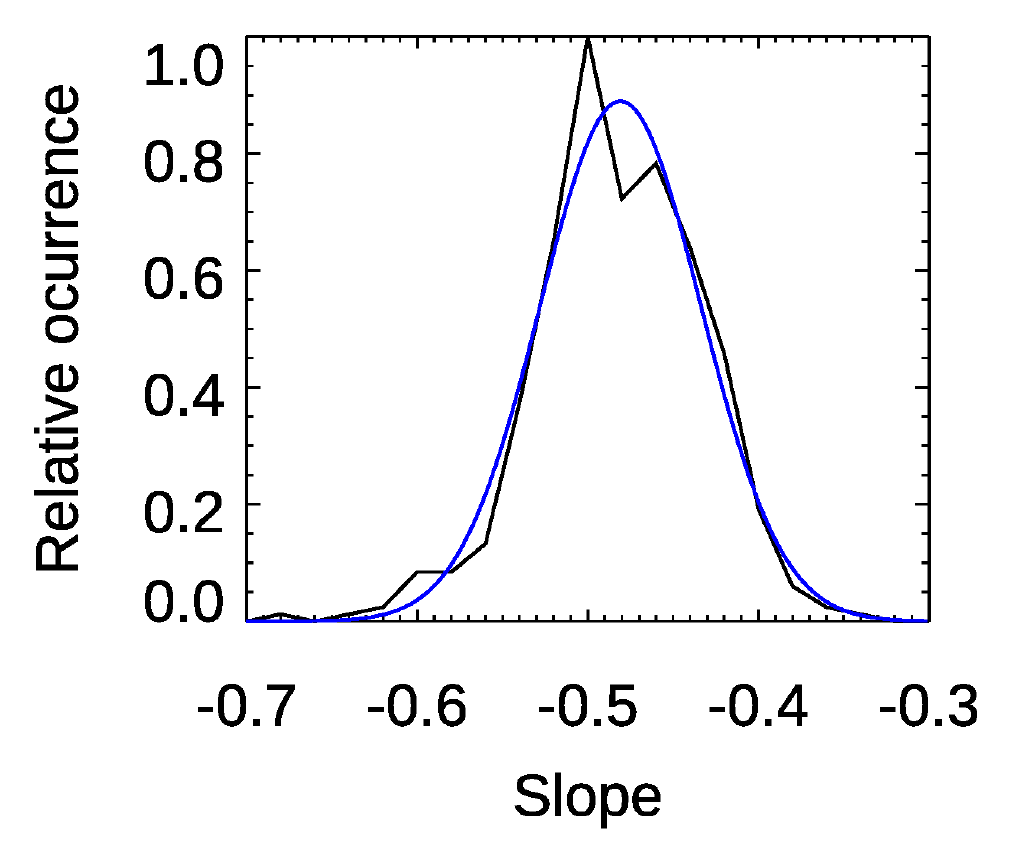}
\caption{Histogram of Allan deviation slopes for the residuals (data minus fit) of 439 eclipses (3.6- and 4.5\,$\mu$m both included).  The blue curve is the best-fitting Gaussian that represents the distribution of slopes, and it peaks at -0.48, with standard deviation 0.047. \label{fig: slopes}}
\end{SCfigure}

The eclipse fit begins with a search over phase to find an initial starting value for center of eclipse.  With the best central phase found, the code then explores many combinations of photometric aperture and bin size, choosing the combination that produces the best slope in the Allan deviation relation at that trial value of the central phase.  The Allan deviation relation \citep{Allan_1966} expresses that the standard deviation of the residuals (data minus best fit) should decrease as the square root of the bin size.  The fitting process at this stage is a multi-variate linear regression.  We bin both the PLD pixel coefficients and the photometry for the fit.  The reasons for binning the data are explained below.  We limit the bin size so that the number of data points after fitting exceeds the number of fitted parameters by at least a factor of 10.  In our experience, this avoids overfitting.  A histogram of the Allan variance slope is shown as Figure~\ref{fig: slopes} for all eclipses we analyze (both 3.6- and 4.5\,$\mu$m combined).  The distribution of slopes is peaked at -0.48, and the distribution closely approximates a Gaussian.  A Gaussian fit to the histogram (illustrated on Figure~\ref{fig: slopes}) has a standard deviation of 0.047, which reflects both the random uncertainty of each slope measurement as well as intrinsic variation in the quality of the data.  Only one eclipse produces a fitted slope that is $3\sigma$ less than -0.5 (AOR 39446016, for WASP-35).  We verified that the seemingly unphysical slope in that instance is not due to overfitting, but occurs because of autocorrelation in the data.

When the fit has selected the best combination of bin size and photometric aperture, it then refines the eclipse depth, central phase, and ramp parameters by exploring parameter space using a Markov Chain Monte Carlo (MCMC) sampling process.  The MCMC is a classic Metropolis-Hastings algorithm with Gibbs sampling.  During a 10,000 step burn-in process, we dynamically adjust the step size to achieve an acceptance ratio of 0.35.  Then each MCMC chain runs for 800,000 steps, and we find that the sampling of parameter space is excellent in both eclipse depth and central phase.

\subsection{Binned Data}
Fitting to binned data is (in theory) non-optimal because the binning process destroys information \citep{Kipping_2010}.  Recently, \citet{May_2021} concluded against using binned data in fitting to phase curves (but not secondary eclipses {\it per se}).  We therefore re-visited our rationale for using binned fits, and we conclude that the binning is desirable, at least in our method.  We here explain our reasons.  First, we note that the {\it Spitzer} IRAC instrument already bins the incident photons in time before transmitting the data to Earth.  The observing cadence available in IRAC was set before transiting planets were discovered, so the observing cadences are not special for the purpose of secondary eclipses.  We find that further binning of the data eliminates instrumental effects on short time scales \citep{Challener_2021} that would otherwise be very difficult to adequately reproduce in the fitting process.  In other words, although the binning destroys information, it destroys information that we want to destroy because we cannot reproduce those short-term instrument effects, and they occur on a faster time scale than the eclipse.  We find that the binning allows the fitting process to concentrate on the time scales that are most relevant to the eclipse, and thereby it reduces red noise.  We limit the size of the bins so that we do not distort the shape of the eclipse curve (e.g, by broadening the ingress/egress), and (as noted above) we limit the bin size so as not to overfit the data.  For planets with multiple observed eclipses, our code usually selects different bin sizes from eclipse to eclipse, because they are independent events.  In those cases, we checked to verify that eclipse depths do not correlate with the adopted bin size.

\subsection{Temporal Baselines}
We initially fit each eclipse with two different temporal baselines (ramps), a linear and quadratic ramp.  We consult a Bayesian Information Criterion (BIC) analysis applied to the fitting residuals in order to decide between the linear and quadratic ramp.  Although the BIC is very useful, we do not use it rigidly, and we also consider the totality of fit properties when deciding between temporal ramps.  For example, the BIC may prefer a quadratic ramp, but that might produce a two-peaked posterior distribution for the central phase, that is not physically realistic.  So in that case, we would adopt the linear ramp.  For planets whose equilibrium temperatures exceed $2000$\,K, we use only a quadratic temporal ramp.  The reason for that choice is to allow the ramp to account for the small portion of the phase curve that occurs within our data window, and thereby avoid bias in the fitted eclipse depth.  We verified that a quadratic is a close approximation to the portion of the sinusoidal phase curve that occurs in our window.

\subsection{Chain Convergence}
After deciding on the ramp, we run an additional MCMC chain with the adopted ramp, and we inspect the posterior distributions for eclipse depth and central phase.  The distributions for two independent chains should be, and are in almost all cases, closely coincident.  We also calculate a Gelman-Rubin (GR) statistic \citep{Gelman_1992} for eclipse depths in the two adopted MCMC chains.  \citet{Gelman_1992} suggested that a value less than 1.1 denotes good convergence, but more recent work has tended to emphasize GR values closer to unity.  The GR statistic is derived from the shape and mean value of the posterior distributions, so closely equal distributions will always produce GR values near unity. Considering all planets and all eclipses, our median GR value is 1.002, our average value is 1.003, and and our maximum value is 1.05.  

Given the large number of eclipses we analyzed, we did encounter some where convergence was initially problematic.  Those cases usually arise when the temporal ramp coefficients become partially degenerate with the eclipse depth. That situation is more common when there is minimal out-of-eclipse baseline before and/or after the eclipse.  In those cases, we achieve good convergence by freezing the temporal ramp coefficients during the MCMC.  Because that freeze can cause the uncertainty in eclipse depth to be underestimated, in those cases we compare the width of the posterior distributions for eclipse depth (that determine the random error) with and without the freeze, and we thereby apply a correction to the error bars.

\subsection{Choice of Eclipse Depth}
Our analysis yields two eclipse depths for each planet; each of those depths is based on average from the same two (or more) converged MCMC chains.  The first depth is the best fit using the criterion adopted by G20, who adopted the solution yielding the best Allan deviation relation.  With our larger sample, we have come to prefer the (more conventional) practice of using the mean of the posterior distribution for depth.  We verified that the conclusions of this paper do not change significantly if we adopt the choice used by G20.

\begin{figure*}
\centering
\includegraphics[width=7in]{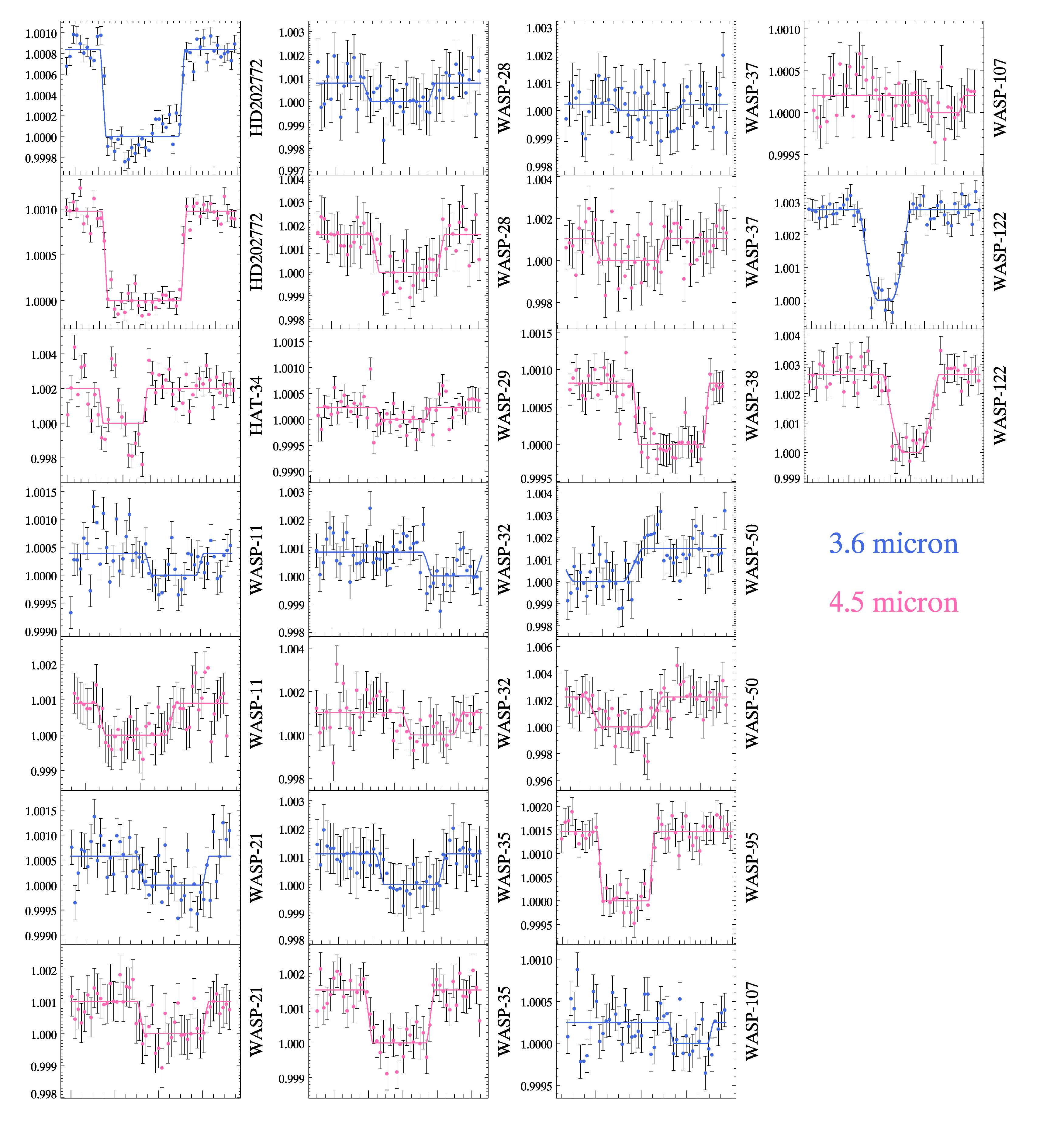}
\caption{Eclipses of planets not previously published, and including WASP-29b that was only recently published by \citet{Wong_2022}. In these plots, the data have been binned, and binning scales are adopted so that the data for each eclipse plot are spanned by 50 plotted points.  The 50-point binning is only for uniformity of illustration, the actual binning used in fitting for the eclipse depths is different in all cases.  The X-axis of the plots is orbital phase, but values are not listed since the central phases of the eclipses are not analyzed in this paper. }  \label{fig: new_eclipses}
\end{figure*}

Our derived eclipse depths and uncertainties are given in Table~\ref{table:depths}. Eclipses for planets not previously published are illustrated in Figure~\ref{fig: new_eclipses}.

\subsection{Special Cases}
There are several planets that represent special cases.  WASP-33b orbits a pulsating star, and \citet{Deming_2012} used a wavelet technique to account for the stellar pulsations while simultaneously solving for the eclipses.  The wavelet technique is not easily included in our PLD code, so we have not re-analyzed the eclipses of WASP-33b, but we include the depths from \citet{Deming_2012} and \citet{Zhang_2018} in Table~\ref{table:depths} (so that all of {\it Spitzer's} eclipse depths at these wavelengths are available in a single Table).  Also, some planets have very weak eclipses observed multiple times, and we treated those slightly differently.  In those cases, we used the PLD code to correct for, and remove, {\it Spitzer's} instrumental effect.  We then combined the photometry for the multiple instrument-corrected eclipses and fitted an eclipse curve. Those fits obtained the eclipse depth and uncertainty based on the scatter, while constraining the central phase to equal 0.5 exactly.  Those cases are for planets WASP-29b and HAT-P-15b in both bands, and WASP-67b and HAT-P-15b in the 3.6\,$\mu$m band.  The depths are listed in Table~\ref{table:depths} associated with the first listed AOR, but all of the listed AORs were averaged to derive the quoted depth.  

\section{Dilution Corrections}\label{sec: dilution}
When a second star appears in the photometry aperture, the light from that companion star will dilute the eclipse, regardless of whether the companion star is physically bound to the planet-hosting star. We corrected the emergent fluxes for dilution using the multiplicative factors in Table~\ref{table:dilute}. The eclipse depths in Table~1 are "as observed", not corrected for dilution.  The dilution factors were applied to the observed eclipse depths to derive the emergent fluxes in Table~\ref{table:fluxes}.

We adopted dilution factors from \citet{Shporer_2014} and G20, and we calculated factors for planets not previously corrected for dilution.  In many cases, published {\it Spitzer} eclipse depths have not been previously corrected for dilution because the diluting companion star was not detected when the {\it Spitzer} eclipses were first analyzed.  We consulted the literature of high resolution imaging to find stellar companions \citep{Ginski_2013, Ngo_2015, Ngo_2016, Evans_2018a, Evans_2018b, Wollert_2015a, Wollert_2015b}.  We also found companions that were sufficiently prominent, and spatially resolved from the primary star, using astronomical catalogs such as 2MASS. Upon identifying a companion, we used the spectral types and K-band magnitudes of both stars to calculate their fluxes in the 3.6- and 4.5\,$\mu$m {\it Spitzer} bands, based on the STAR-PET tool\footnote{https://irsa.ipac.caltech.edu/data/Spitzer/docs/dataanalysistools/tools/pet/starpet/}.  Companions closer than $\sim 2\,arc-sec$ were not resolved from the primary star, and their fluxes simply add to produce the dilution. When the companion star was spatially resolved in the {\it Spitzer} data, we used the method described by G20 (their Section~3.5) to calculate the fraction of its flux that was scattered into the photometric aperture centered on the primary star.  Using the calculated fluxes from both stars in the photometric aperture, we calculated the dilution correction factors at 3.6- and 4.5\,$\mu$m listed in Table~\ref{table:dilute}.   

\section{Comparison to Literature Values}

\begin{figure}
\centering
\includegraphics[width=6in]{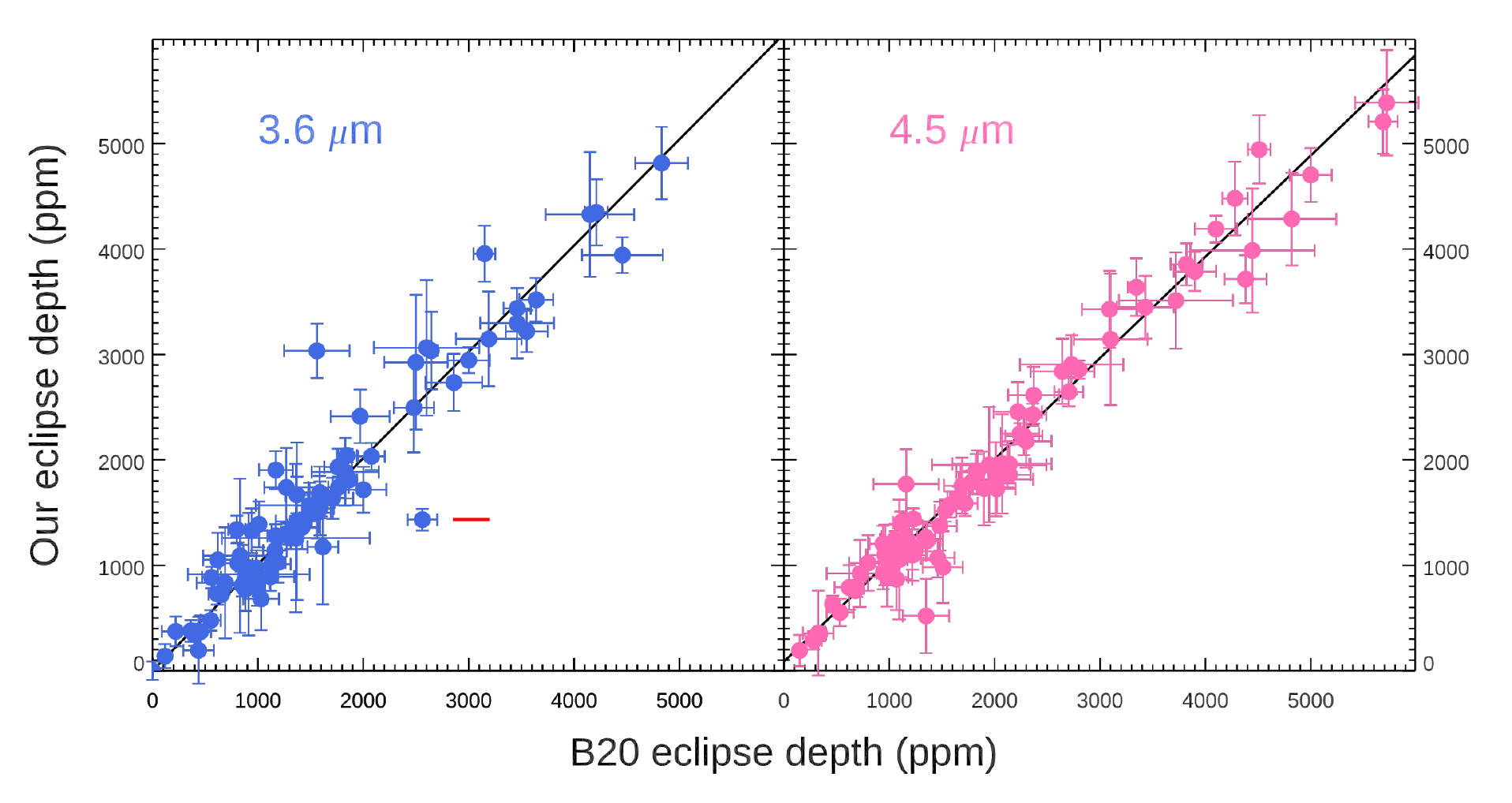}
\caption{Average eclipse depth per planet derived in this paper, versus literature values compiled by \citet{Baxter_2020}. The black lines are least-squares straight lines, and their slopes are within $2\sigma$ of unity, being $0.961\pm0.020$ and $1.008\pm0.025$ at 3.6- and 4.5\,$\mu$m, respectively. The red mark on the 3.6\,$\mu$m panel marks the outlying point for HD\,189733b - see text in Appendix Section C for discussion. \label{fig: compare1}}
\end{figure}

Although our eclipse depths are more extensive than any extant eclipse data set, it is nevertheless interesting to compare subsets of our values to previous investigations.  Previous compilations of eclipse depths have been presented by G20, B20, W20, and \citet{Dransfield_2020}.  We choose to compare to the values compiled by B20, especially because those authors found an abrupt increase in the ratio of 4.5- to 3.6\,$\mu$m eclipse depth - a phenomenon similar to the abrupt flux increase that we infer in this paper.  Figure~\ref{fig: compare1} shows our 3.6- and 4.5\,$\mu$m average eclipse depths per planet, as a function of the depth from B20, for planets in common. 

Least-squares fits of straight lines to Figure~\ref{fig: compare1}, considering the uncertainties in both X and Y, yield slopes within $2\sigma$ of unity. We conclude that there is no systematic scale factor difference between our new eclipse depths and literature values.  We also investigated outlying points on Figure~\ref{fig: compare1}.  The most prominent outlier is HD\,189733b at 3.6\,$\mu$m (marked on Figure~\ref{fig: compare1}), where our average eclipse depth for nine eclipses is $1432\pm34$ ppm (median value $1467$ ppm).  B20 list $2560\pm140$, based on one eclipse from \citet{Charbonneau_2008}, so this planet lies distinctly to the right of the fitted line in the left panel of Figure~\ref{fig: compare1}.  Consequently, the 3.6\,$\mu$m brightness temperature calculated by B20 ($1604\pm32$ Kelvins), is well above the equilibrium temperature of 1191 Kelvins.  Our 3.6\,$\mu$m brightness temperature is $1337\pm30$ Kelvins. Our values are consistent with a re-analysis by \citet{Knutson_2012}, who found eclipse depths consistent with our current results. Given that most planets at this level of irradiation have $T_b$ not greatly exceeding their equilibrium temperatures, and given that our nine eclipses are mutually consistent, we confirm the result from \citet{Knutson_2012} that the original 3.6\,$\mu$m eclipse value from \citet{Charbonneau_2008} is in need of revision, mostly likely due to improvements in observing and data analysis techniques.  We point out that a similar revision for HD\,209458b was found by \citet{Diamond-Lowe_2014}, and we note that our eclipse depths for HD\,209458b are also in good agreement with that revision.

\begin{figure}
\centering
\includegraphics[width=6in]{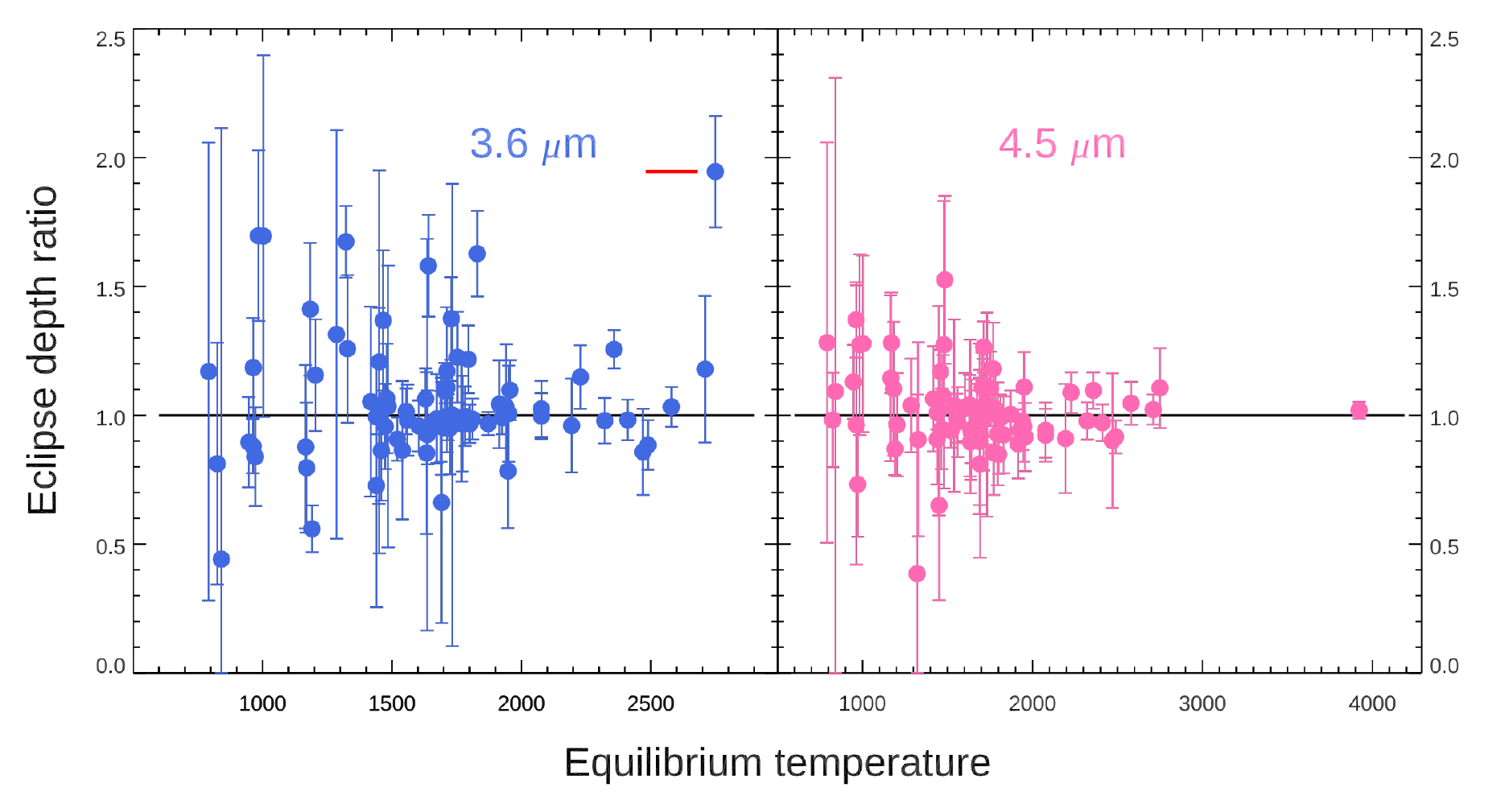}
\caption{Ratio of the average eclipse depth per planet (ours divided by \citealp{Baxter_2020}), versus planet equilibrium temperature.  We find no systematic variations at either wavelength as a function of equilibrium temperature. The red mark indicates the outlier Kepler-13b, discussed in Section C of the Appendix. \label{fig: compare2}}
\end{figure}

We also compared the {\it ratio} of our eclipse depth to the value from B20, as a function of equilibrium temperature.  That comparison is shown in Figure~\ref{fig: compare2}, and in this case also we have investigated a prominent outlier at 3.6\,$\mu$m. The outlier (red mark on Figure~\ref{fig: compare2}) is Kepler-13b, where B20 use an eclipse depth of $1560\pm330$\,ppm, from \citet{Shporer_2014}.  Our eclipse depth ($3035\pm257$\,ppm) is almost twice as large.  This difficult system has a bright companion star separated by approximately 1 pixel, that produces a large dilution correction and adds noise to the photometry via pointing jitter.  We use the same dilution correction as do \citet{Shporer_2014}, and we think that our PLD analysis method is well matched to correct for noise produced by the companion star. The relatively "small" eclipse depths derived by \citet{Shporer_2014} lead them to infer an "efficient heat redistribution process" that is unusual for strongly irradiated hot Jupiters \citep{Cowan_2011}, and they also derive a geometric albedo (0.33) at the high end of the range for hot Jupiters.  We have rechecked our eclipse data analyses for this planet: we find no issues that need correction, and we believe that our eclipse depths are correct, albeit with large random errors.  Pending further analysis of other data on this system, we tentatively suggest that our eclipse depths are more realistic than are the values derived by \citet{Shporer_2014}, and used by B20.  However, we also note that the population-level results in this paper are insensitive to the Kepler-13 results.  

\section{PHOENIX versus ATLAS Models}

PHOENIX stellar atmospheres \citep{Hauschildt_1997} are used in our planetary modeling code (Section~\ref{sec: models}), whereas we use ATLAS9 stellar models to calculate emergent planetary fluxes from eclipse depths (Section~\ref{sec: derivation}). These models are respectively deeply rooted in our procedures, and it would be impractical to alter those choices. Hence, we investigated to what degree different stellar models can affect our results. The principal concern was to what degree the choice of model type affected the emergent day side flux that we calculated for each planet (Section~\ref{sec: derivation}).  B20 investigated PHOENIX versus ATLAS models and noted that planetary brightness temperatures were consistent between PHOENIX and ATLAS.  \citet{Husser_2013} compared stellar fluxes from PHOENIX and ATLAS and found excellent agreement.  We investigated further by calculating stellar fluxes in the 3.6- and 4.5\,$\mu$m Spitzer bands, from both PHOENIX (BT-NextGen) and ATLAS9.  We calculated the ratio of ATLAS flux to PHOENIX flux at 9 stellar temperatures from 4000K to 9000K, at $\log{g}=4.5$.  The ratio differed from unity with an average absolute difference of $0.68$\% and $1.02$\% at 3.6- and 4.5\,$\mu$m, respectively.  Those differences are sufficiently small to have no significant impact on our results. Only at 9000 Kelvins did we find a significant difference, where the ATLAS9 stellar flux is $4$\% less than from PHOENIX.  That would affect primarily KELT-9b, and to a lesser extent KELT-20b.  However, \citet{Witzke_2021} found that PHOENIX models underestimate ultraviolet flux for hot stars compared to observations (their Figure~17).  Because the models constrain the emergent stellar bolometric flux to be closely equal to $\sigma T_{eff}^4$, underestimating UV flux will require overestimating emergent flux at other wavelengths (e.g., the Spitzer bands).  Consequently, we prefer the ATLAS9 models over PHOENIX especially for the hottest stars, to avoid that issue.

\clearpage
\begin{table*}[t]
\centering\caption{Eclipse depths and uncertainties for 3.6- and 4.5\,$\mu$m  {\it Spitzer} observations, in units of the stellar flux, in parts-per-million. All planets are the 'b' planet in the system unless otherwise noted.  These are "as observed" eclipse depths, not corrected for dilution by companion stars.  AOR = Astronomical Observing Request number.  In the rare instances where more than one eclipse occurs in a single AOR, the depths are listed in chronological order. For four planets (HAT-P-12 and -15, WASP-29 and -67), several AORs were averaged to yield a single eclipse depth. These eclipse depths are converted to emergent fluxes and brightness temperatures in Table~\ref{table:fluxes}.  (This Table is published in its entirety in machine-readable format. A portion is shown here for guidance regarding its form and content.) }
\begin{tabular}{llrrl}
Planet & $\lambda$ ($\mu$m) & AOR & Depth & $1\sigma$  \\
\hline
\hline
CoRoT-01 &  3.6  &      36786176 &      4328 &       591 \\ 
CoRoT-01 &  4.5 &      36785920 &      4285 &       440 \\ 
CoRoT-02 &  3.6  &      31774976 &      3134 &       198 \\ 
CoRoT-02 &  4.5 &      28640000 &      4788 &       272 \\ 
      &   &     57958144  &      4660 &       308 \\
      &   &     57958656  &      4310 &       369 \\
HAT-P-01 &  3.6  &       2474547 &      1339 &       129 \\ 
HAT-P-01 &  4.5 &      21060608 &       520 &       351 \\ 
\end{tabular}
\label{table:depths}
\end{table*}

\begin{longtable}{lllllll} 

\caption{Emergent fluxes in the 3.6- and 4.5\,$\mu$m bands, and brightness temperatures in Kelvins. Planets are the b planet in each system. Units of fluxes are ergs cm$^{-2}$ sec$^{-1}$ Hz$^{-1}$, or (equivalently) milli-Watts m$^{-2}$ Hz$^{-1}$. (This Table is published in its entirety in machine-readable format. A portion is shown here for guidance regarding its form and content.) } \\

Planet & F$_{\nu}$ 3.6\,$\mu$m & $1\sigma$ & F$_{\nu}$ 4.5\,$\mu$m & $1\sigma$ & T$_b$ 3.6\,$\mu$m &  T$_b$ 4.5\,$\mu$m  \\
\hline
\hline
   CoRoT-01  &   6.42e-06  &   8.76e-07  &   4.01e-06  &   4.12e-07  &  $2413^{+ 156}_{- 161}$  &  $2145^{+ 112}_{- 115}$  \\
    CoRoT-02  &   3.20e-06  &   1.97e-07  &   2.95e-06  &   1.98e-07  &  $1785^{+  43}_{-  44}$  &  $1844^{+  58}_{-  59}$  \\
    HAT-P-01  &   2.97e-06  &   2.87e-07  &   7.26e-07  &   4.90e-07  &  $1733^{+  64}_{-  65}$  &  $1073^{+ 206}_{- 298}$  \\
    HAT-P-02  &   4.58e-06  &   9.66e-07  &   3.68e-06  &   4.56e-07  &  $2067^{+ 185}_{- 196}$  &  $2052^{+ 127}_{- 130}$  \\
    HAT-P-03  &   1.87e-06  &   2.83e-07  &   1.55e-06  &   2.20e-07  &  $1467^{+  73}_{-  77}$  &  $1401^{+  75}_{-  80}$  \\
    HAT-P-04  &   5.51e-06  &   8.30e-07  &   2.81e-06  &   6.25e-07  &  $2247^{+ 153}_{- 159}$  &  $1803^{+ 182}_{- 192}$  \\
  
\label{table:fluxes}
   \end{longtable}

\begin{longtable}{lrrrr}
\caption{Stellar and planetary temperatures and radii that we adopted from TEPCat. Planets are the 'b' planet in each system. The stellar radii are in solar units, and the planetary radii in Jupiter units.  Stellar temperatures are effective temperatures.  The planetary temperatures are the calculated equilibrium temperatures based on the stellar irradiance, and adopting an albedo of zero and uniform redistribution over the planetary sphere.  The values listed for HAT-P-2 and HD\,80606 are not from TEPCat, they are the day side temperatures from the time-dependent calculations of \citet{mayorga_2021}. (This Table is published in its entirety in machine-readable format. A portion is shown here for guidance regarding its form and content.) }\\

Planet & $R_{planet}$ & $T_{eq}$ & $R_{star}$ & $T_{star}$  \\
\hline
\hline
    CoRoT-01  &   1.551  &    1915  &   1.131  &    5950  \\  
    CoRoT-02  &   1.460  &    1521  &   0.901  &    5598  \\  
    HAT-P-01  &   1.319  &    1322  &   1.174  &    5975  \\  
\label{table: parameters}        
\end{longtable}

\begin{table}
\centering\caption{Dilution corrections for secondary eclipse depth at both {\it Spitzer} wavelengths. Sources: G20 = Garhart et al.(2020), TH = this paper, SH = Shporer et al.(2014). }
\begin{tabular}{llll}
Planet & 3.6\,$\mu$m factor & 4.5\,$\mu$m factor & Source \\
\hline
\hline
CoRoT-2b  &  1.0276  & 1.0255 &  TH  \\
HAT-P-8b  &  1.0109  & 1.0102 &  TH \\
HAT-P-16b &  1.0006  & 1.0006 &  TH \\
HAT-P-20b &  1.0070  & 1.0074 &  TH \\
HAT-P-30b &  1.0121  & 1.0117 & G20 \\
HAT-P-33b &  1.0377  & 1.0332 & G20 \\
HAT-P-41b &  1.0069  & 1.0111 & G20 \\
HD\,202772b & 1.207  & 1.207  & TH  \\
KELT-1b   &  1.0065  & 1.0059 & TH  \\
KELT-2b  &   1.137   & 1.123  & G20 \\ 
KELT-3b   &  1.0125  & 1.0127 & G20 \\
KELT-16b  &  1.0189  & 1.0168 & TH   \\
Kepler-5b &  1.0434  & 1.0384 & TH   \\
Kepler-6b &  1.0041  & 1.0049 & TH   \\
Kepler-13b & 1.9940  & 1.9942 & SH   \\
TrES-2b   &  1.0812  & 1.0741 & TH   \\
TrES-4b   &  1.0258  & 1.0242 & TH  \\
WASP-1b   &  1.0015  & 1.0020 & TH  \\
WASP-2b   &  1.0894  & 1.0870 & TH  \\
WASP-8b   &  1.1058  & 1.0977 & TH  \\
WASP-11b  &  1.0630  & 1.0720 & TH  \\
WASP-12b  &  1.1101  & 1.0983 & G20 \\
WASP-14b  &  1.0126  & 1.0115 & TH  \\
WASP-26b  &  1.0601  & 1.0690 & TH  \\
WASP-33b  &  1.0038  & 1.0035 & TH  \\
WASP-36b  &  1.0015  & 1.0026 & G20 \\
WASP-49b  &  1.0130  & 1.0124 & G20 \\
WASP-72b  &  1.0484  & 1.0411 & TH  \\ 
WASP-76b  &  1.1470  & 1.1250 & G20 \\ 
WASP-77b  &  1.0929  & 1.0530 & G20 \\
WASP-87b  &  1.0014  & 1.0011 & G20 \\
WASP-103b &  1.1700  & 1.1490 & G20 \\
WASP-131b &  1.0781  & 1.0663 & TH   \\
\end{tabular}
\label{table:dilute}
\end{table}


\clearpage

\bibliography{references.bib}
\end{document}